%% file: los_slacs_ii_accepted.tex
\documentclass[twocolumn]{openjournal}

\input{packages}

\begin{document}
\title{Line-of-sight shear in \emph{SLACS} strong lenses II: \\
validation tests with an extended sample}

\author{Natalie B. Hogg$^{1, 2, 3, \ast}$,  Daniel P. Johnson$^4$, Anowar J. Shajib$^{5, 6, 7}$, Julien Larena$^4$}
\thanks{$^\ast$natalie.hogg@ast.cam.ac.uk}

\affiliation{$^1$ Institute of Astronomy,  University of Cambridge, Madingley Road, Cambridge, CB3 0HA, UK}
\affiliation{$^2$ Kavli Institute for Cosmology, University of Cambridge, Cambridge, UK}
\affiliation{$^3$ Clare Hall, Herschel Road, Cambridge, CB3 9AL, UK}
\affiliation{$^4$Laboratoire Univers et Particules de Montpellier, CNRS \& Université de Montpellier,\\Parvis Alexander Grothendieck, Montpellier, France, 34090}
\affiliation{$^5$Department of Astronomy \& Astrophysics, University of Chicago, Chicago, IL 60637, USA}
\affiliation{$^6$Kavli Institute for Cosmological Physics, University of Chicago, Chicago, IL 60637, USA}
\affiliation{$^7$Center for Astronomy, Space Science and Astrophysics, Independent University, Bangladesh, Dhaka 1229, Bangladesh}

\begin{abstract}
Strong gravitational lensing images are subject to shape distortions due to inhomogeneities along the line of sight. The leading order shape distortion is shear, which, if measurable, will be a complementary cosmological probe to traditional cosmic shear. In \cite{Hogg:2025wac}, we modelled 23 of the SLACS strong lenses, studying the line-of-sight (LOS) shear under a variety of shear and mass model parametrisations. In this work, we successfully model 22 of an additional 27 lenses, extending our sample of LOS shear constraints to 45 in total. We find a mean shear magnitude of $0.11\pm 0.024$, showing that a significant fraction of the lenses modelled in this work possess LOS shears with unexpectedly large magnitudes, $|\gamma_{\rm LOS}| > 0.1$, even when an octupolar distortion is included in the lens mass. We further investigate if factors such as lens and source redshift, filter and PSF, or flux and signal-to-noise ratio in the lensed arcs correlate with shear. We find that none of these features play a statistically significant role in the production of unusually large shear magnitudes.
\end{abstract}


\section{Introduction}

Every beam of light propagating in the Universe experiences gravitational lensing, the deflection of its path by some mass that has induced local curvature in spacetime. These deflections may be small, and undetectable when observing single sources -- the regime of \textit{weak} lensing -- or, given good alignment of a sufficiently concentrated mass, such as a massive elliptical galaxy, with a source, the deflections may be large, producing multiple images of that source. This is the regime of \textit{strong} lensing. Measuring the effects of gravitational lensing thus provides a direct probe of the behaviour and distribution of dark matter on scales ranging from individual galaxies to the large-scale structure in the Universe.

Weak and strong gravitational lensing are rich and mature fields of study, each serving as a precision tool for cosmological inquiry \citep{Shajib2024, Prat:2025ucy}. In recent years, increasing attention has been given to the intersection between these domains -- the impact of weak lensing on strong lensing -- both as a potential source of bias (see e.g. \cite{Jaroszynski2014,McCully2017,Johnson:2024hvl}), but also as a novel cosmological probe in its own right \citep{Birrer:2016xku, Birrer:2017sge, Fleury:2021tke}.

In the following, we briefly recap gravitational lensing formalism (see e.g. \cite{Kilbinger:2014cea} for a complete review), and how weak lensing contributions from the line of sight affect strong lensing observables. We summarise the motivation for and status of efforts to constrain these effects, and, following on from \cite{Hogg:2025wac}, present the latest results and meta-analysis from a study of line-of-sight distortions in real strong lensing images from the SLACS catalogue \citep{Bolton2006, Bolton2008}.
\subsection{Gravitational lensing formalism}
The equation which relates an unlensed source position, $\bm{\beta}$, to the observed lensed image position $\bm{\theta}$, is called the lens equation,
\begin{equation}
    \bm{\beta} = \bm{\theta} - \bm{\alpha}(\bm{\theta}), \label{eq:lenseqn}
\end{equation}
where the displacement angle $\bm{\alpha}(\bm{\theta})$ is related to the gravitational potential of the deflector projected along the line of sight into a single plane,
\begin{equation}
	\bm{\alpha}(\bm{\theta})
	= \ddf{\psi(\bm{\theta})}{\bm{\theta}},
\end{equation}
and where
\begin{equation}
	\psi(\bm{\theta})
	\equiv
	\frac{D\e{ds}}{D\e{od}D\e{os}} \, \hat{\psi}(D\e{od}\bm{\theta}) \ ,
\end{equation}
$\hat{\psi}(\bm{x})$ being twice the projected gravitational potential produced by the surface density of the main lens, $\Sigma(\bm{x})$,
\begin{equation}
	\hat{\psi} (\bm{x})
	\equiv
	\frac{4G}{c^2} \int \dd^2 \bm{y} \; \Sigma(\bm{y}) \ln |\bm{x} - \bm{y}| \ ,
\end{equation}
where $\bm{x}$ is the vector describing the point where the light ray strikes the lens plane, $G$ is Newton's constant and $c$ is the speed of light.

\subsection{Weak lensing}

When $\bm{\alpha}(\bm{\theta})$ is approximately constant over the angular extent of the source, this only gives rise to distortions of the image, without forming multiple images. This is usually the case when the lensing object is diffuse or far from the line of sight. This situation is well-approximated by a local linearisation of the lens equation,
\begin{equation}
    \delta \bm{\beta} \approx \bm{\mathcal{A}} \; \delta \bm{\theta},
\end{equation}
where the lensing Jacobian $\bm{\mathcal{A}}$ gives the mapping between the image and source plane,
\begin{equation}
        \mathcal{A}_{ij} = \frac{\partial\beta_i}{\partial\theta_j} = \delta_{ij} - \frac{\partial \alpha_i}{\partial \theta_j}.\label{eq:Jacobian}
\end{equation}

This is typically decomposed as 
\begin{equation}
    \bm{\mathcal{A}} = \begin{bmatrix}
  1 - \kappa - \rm{Re}\left(\gamma\right) & -\rm{Im}\left(\gamma\right) \\
  -\rm{Im}\left(\gamma\right) & 1 - \kappa + \rm{Re}\left(\gamma\right)
  \end{bmatrix}, \label{eq:Amatrix}
\end{equation}
and is taken to be constant on the angular scales of individual weakly lensed galaxies. $\kappa$ is called the convergence and $\gamma = \gamma_1 + i\gamma_2$ is called the shear. Convergence describes the isotropic scaling of image size, whilst shear describes the anisotropic change of image shape and can be equivalently parameterised as $\gamma = |\gamma|\exp{(2i\varphi)}$, where $\varphi$ is the position angle of the shape distortion around some pre-defined axis. 

Shear measurements are made by observing the shapes of objects, typically galaxies, but also, as is the focus of this paper, strong gravitational lenses.\footnote{The mass-sheet degeneracy \citep{Falco1985, Schneider:2013wga} renders convergence unmeasurable, a consequence of which is that shear measurements are in fact only sensitive to the `reduced' shear, $g = \gamma/(1-\kappa)$. However, given that both $\kappa$ and $\gamma$ are typically very small, $g$ and  $\gamma$ can be treated as being approximately equivalent.} Measuring the shear from individual images of galaxies is inherently noisy, due to their intrinsic elliptical shapes, and observational effects such as pixelisation and blurring due to the point spread function (PSF), and atmosphere in the case of ground-based telescopes. 

These challenges have been overcome by leveraging large galaxy surveys from which millions or billions of shape measurements can be made, yielding sufficient signal over noise \citep{Bacon:2000sy, Kaiser:2000if, vanWaerbeke2000, Wittman:2000tc, Lin2012, Kilbinger2013, Kuijken:2015vca, Secco2021}; or alternatively by exploiting ultra-high resolution space-based imaging of smaller areas of the sky, for example with the James Webb Space Telescope \citep{Scognamiglio:2026phv}. The two-point correlation function of galaxy shapes can then be estimated -- also known as the `cosmic shear' signal -- and cosmological constraints derived \citep{DES:2021bvc, DES:2021vln, DECV2025}.

\subsection{Shear from strong lenses}
As an alternative to estimating shear from galaxy shapes, strong gravitational lensing images may be used instead \citep{Birrer:2016xku, Birrer:2017sge, Fleury:2021tke}. If shape distortions due to weak lensing are measurable in strong lensing images, this will provide complementary constraints on cosmology to those of weak lensing measured in galaxy surveys, whilst also avoiding the systematic biases possible in those surveys, such as shape noise and intrinsic alignments.

Even if shear measurements are not the goal, the effect of shear is included in typical strong lens modelling programmes. However, various works have called into question whether this shear is truly `external' --  in other words, physically induced by line-of-sight objects -- or whether it is `internal', acting as a fudge factor to absorb any freedom in the mass model which is not explicitly accounted for \citep{Etherington:2023yyh}.

This distinction has further led to the terminology of `residual' shear being introduced by \cite{Shajib2024},
intended to make explicit that the measured shear might have both internal and line-of-sight contributions.

The final inmate in the nomenclature zoo is `line-of-sight' shear. This term is intended to mean only that shear which is truly `external' to the main lens (though in practice this might not always be distinguishable from `internal' contributions). Furthermore, it has a different sensitivity to mass in the foreground and background of the lens system, and is thus not formally equivalent to the weak lensing shear on the same line of sight. This formulation is intended to fully account for degeneracies from source-position transforms and between the contributions from different components of the line of sight, and allows for a more physical interpretation of line-of-sight effects on strong lenses \citep{Fleury:2021tke}. 

Even when the mass of the main lens is perfectly reconstructed, the line-of-sight shear will be sensitive to any correlated structures in the immediate environment of the lens galaxy which are not explicitly included in the model, as well as less (or non) correlated structures along the line of sight. These choices must therefore be kept in mind when interpreting the resulting shear measurements. If line-of-sight shear measurements are used to study lens environments, this situation is ideal -- if instead the shear is intended as a study of cosmologically representative lines of sight, this will be a bias which must be accounted for.

\subsection{The minimal model for line-of-sight shear}
The majority of strong lens galaxies are early type -- in other words, the distribution of their stellar mass is elliptical. In strong lens modelling, these are often described using an elliptical power law (EPL) potential \citep{Tessore:2015baa}, with a convergence given by
\begin{equation}
    \kappa(x, y) = \frac{3 - \gamma^{\rm EPL}}{2} \left(\frac{\theta_{\rm E}}{\sqrt{q x^2 + y^2/q}}\right)^{\gamma^{\rm EPL}-1}, \label{eq:epl}
\end{equation}
where $\theta_{\rm E}$ is the Einstein radius, $\gamma^{\rm EPL}$ is the slope of the power law describing the three-dimensional mass distribution and $q$ is the axis ratio of the ellipse.

Measuring residual shear via the decomposition shown in \autoref{eq:Amatrix} in conjunction with an elliptical lens mass model presents a challenge, as the ellipticity components of the lens mass will be approximately degenerate with the shear components. In their treatment of line-of-sight (LOS) effects in strong lensing, \cite{Fleury:2021tke} generalised the treatment of LOS perturbations, and were able to construct a \textit{minimal} model for LOS effects, which provides a shear term that is free from degeneracy with the parameters of the main lens model.

In the presence of tidal perturbers on the line of sight,
\autoref{eq:lenseqn} becomes 
\begin{equation}
\bm{\beta} = \amplification\e{os} \bm{\theta} - \amplification\e{ds} \, \frac{ \dd \psi(\amplification\e{od} \bm{\theta})}{\dd \bm{\theta}},
\label{eq:lens_eqn_los}
\end{equation}
where the amplification matrices $\mathcal{\bm{A}}_{\rm os}$, $\mathcal{\bm{A}}_{\rm ds}$ and $\mathcal{\bm{A}}_{\rm od}$ encapsulate the LOS perturbations between observer and source, main deflector and source, and observer and main deflector respectively. These matrices are decomposed in the same way as \autoref{eq:Amatrix}, with an additional rotation term $\omega$ in the trace-free part of the matrix, which arises due to lens--lens coupling,
\begin{equation}
 \bm{\mathcal{A}}_{\rm ab}= \begin{bmatrix}
		1 - \kappa\e{ab} - \rm{Re}\left(\gamma\e{ab}\right) & -\rm{Im}\left(\gamma\e{ab}\right) + \omega\e{ab} \\ -\rm{Im}\left(\gamma\e{ab}\right) - \omega\e{ab} & 1-\kappa\e{ab}+ \rm{Re}\left(\gamma\e{ab}\right)
	\end{bmatrix}.
\end{equation}
As discussed, the individual shear terms in \autoref{eq:lens_eqn_los} are subject to degeneracies with one another, with parameters of the model used to describe the main lens, and with the position of the source. The minimal model of \cite{Fleury:2021tke} re-parametrises the lens equation as
\begin{equation}
\tilde{\bm{\beta}} = \amplification\e{LOS} \bm{\theta} - \frac{\dd \psi_{\rm eff}}{\dd \bm{\theta}} \ , \label{eq:lens_eqn_minimal}
\end{equation}
with the transformed source position $\tilde{\bm{\beta}} \equiv \amplification\e{od}\amplification\e{ds}^{-1}\bm{\beta}$, the LOS amplification matrix $	\amplification\e{LOS} \equiv \amplification\e{od} \amplification\e{ds}^{-1} \amplification\e{os}$ and the effective gravitational potential $\psi\e{eff}(\bm{\theta}) \equiv \psi(\amplification\e{od}\bm{\theta})$. \autoref{eq:lens_eqn_minimal} describes a main lens with potential $\psi\e{eff}$ and external tidal perturbations, $\amplification\e{LOS}$, located in the same plane, and is equivalent to \autoref{eq:lenseqn} up to a source-position transformation. 

It is therefore the LOS shear, the combination $\gamma_{\rm LOS} \equiv \gamma_{\rm od} + \gamma_{\rm os} - \gamma_{\rm ds}$, which is expected to be free from perfect line of sight or mass model degeneracies, and thus carry measurable cosmological information. This latter point was explicitly shown by \cite{Hogg:2022ycw} using mock images of strong lenses. 

In \cite{Hogg:2025wac}, we modelled 23 strong gravitational lenses from the SLACS catalogue using the minimal LOS model, making the first measurements of the LOS shear, $\gamma_{\rm LOS}$. We further showed that seven of these 23 lenses have shear magnitudes at least $3\sigma$ away from the expected values for those lens and source configurations, as computed from an $N$-body simulation. Such large shear magnitudes have been attributed to unmodelled lens mass complexity, notably an octupole, otherwise known as `boxyness' or `diskyness' \citep{Bender1988, Etherington:2023yyh}. Fitting our 23 lenses with a model including an octupole in the lens mass, we found only a single case where the fit improved and the shear magnitude significantly decreased. We therefore concluded that the inclusion of the octupole does not systematically suppress shear magnitudes.

In this companion paper to \cite{Hogg:2025wac}, hereafter referred to as Paper I, we present the results of modelling a further 27 strong gravitational lenses from the SLACS sample using the minimal LOS shear model with the aim of measuring LOS shear in all of these lenses. Following the aforementioned conclusions of Paper I, we focus on data properties unrelated to the lens mass modelling, namely the redshifts of the lenses and sources, and the filters, PSF, flux and signal-to-noise ratio of the images, and explore correlations between these features and our measured shear magnitudes.

\section{Data and methodology}\label{sec:methods}
\cite{Shajib2021} selected 50 strong gravitational lenses from the SLACS catalogue to study the dark matter haloes of elliptical galaxies in a joint lensing--kinematics analysis, of which 23 were successfully modelled. It was from these 23 that we measured LOS shear in Paper I; in the current work we attempt to model the remaining 27 lenses from the initially selected 50, which were themselves chosen from the SLACS catalogue specifically due to their lack of satellite galaxies, their simple source morphologies, and the availability of photometry in the F555W and F606W Hubble Space Telescope (HST) bands, in which the lensed source emission is more easily distinguished from the lens light than F814W. The data were reduced by \cite{Shajib2021} using the \texttt{Astrodrizzle} package \citep{Avila2015}, and the point spread function (PSF) for each filter and camera combination was obtained with  the \texttt{TinyTim} software \citep{Krist2011}.

In modelling this dataset, we follow the same methodology as described in Paper I. In brief, we use the \texttt{dolphin} package \citep{Shajib2021, Tan2024, Shajib:2025bho} to perform semi-automated forward modelling of our strong lens imaging data, applying the same lens models to every image in the sample. The \texttt{dolphin} package uses the \texttt{lenstronomy} software \citep{Birrer:2018xgm, Birrer2021} as its modelling engine.

We model each deflector with an elliptical power law mass profile (\autoref{eq:epl}) and the minimal LOS shear model described above, with a double elliptical S\'ersic profile to model the lens light \citep{Sersic1963,Sersic1968}. The source is modelled by an elliptical S\'ersic and a basis set of Gaussian shapelets \citep{Refregier:2001fd}. The maximum order for the shapelets in each lens model is listed in \autoref{tab:lenses}.

The parameters of the model are first optimised using a particle swarm optimisation \citep{Eberhart1995}, and the best-fit values are used to initialise joint posterior sampling with the affine-invariant ensemble Markov chain Monte Carlo sampler \texttt{emcee} \citep{GoodmanWeare, ForemanMackey2013}. The priors on all model parameters are listed in \autoref{tab:priors}; our results are presented after post-processing the chains, including the removal of burn-in and smoothing of the marginalised posterior distribution histograms.

\section{Results and discussion}\label{sec:results}
\begin{figure*}
	\centering
	\includegraphics[width=\textwidth]{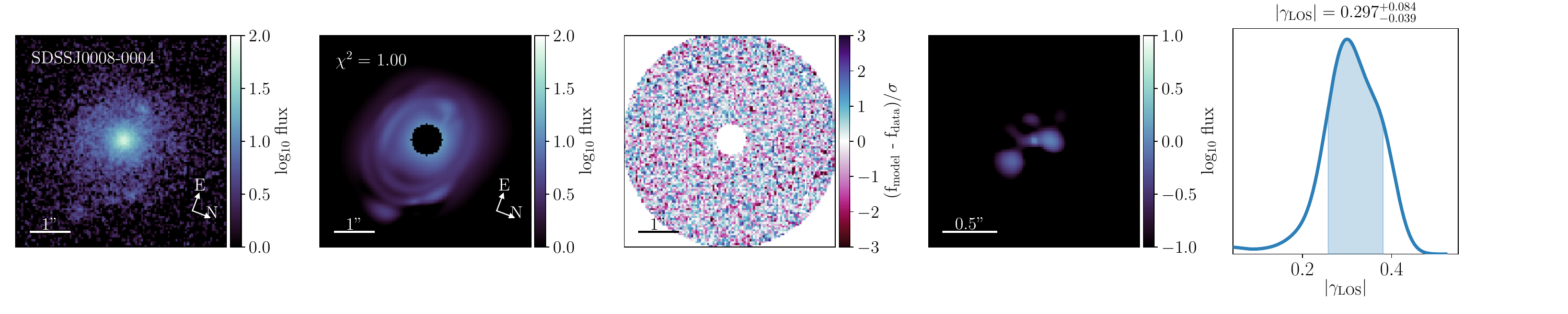}\\
	\includegraphics[width=\textwidth]{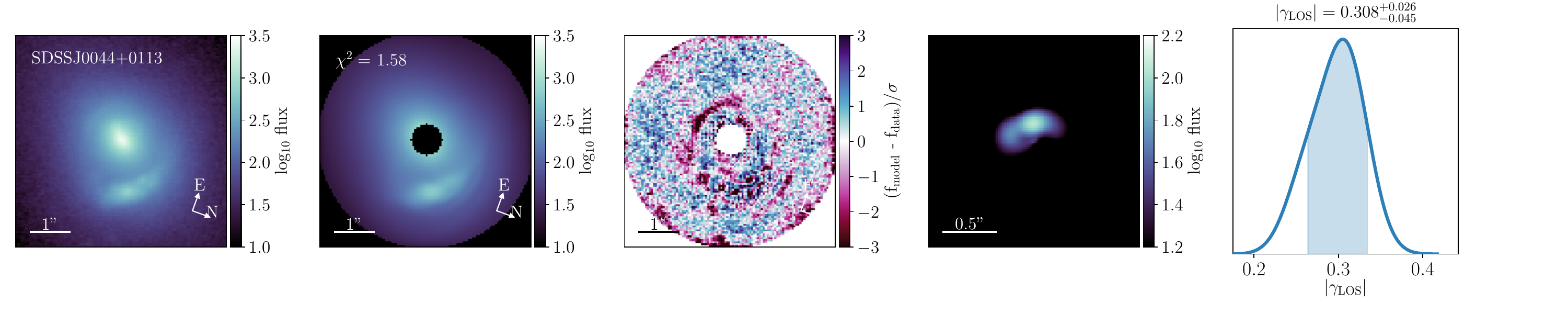}\\
    \includegraphics[width=\textwidth]{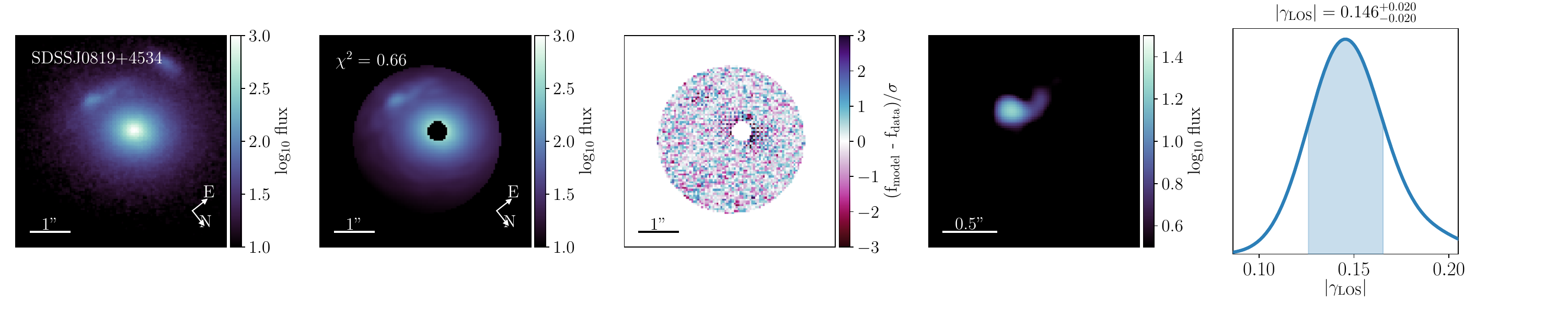}\\
    \includegraphics[width=\textwidth]{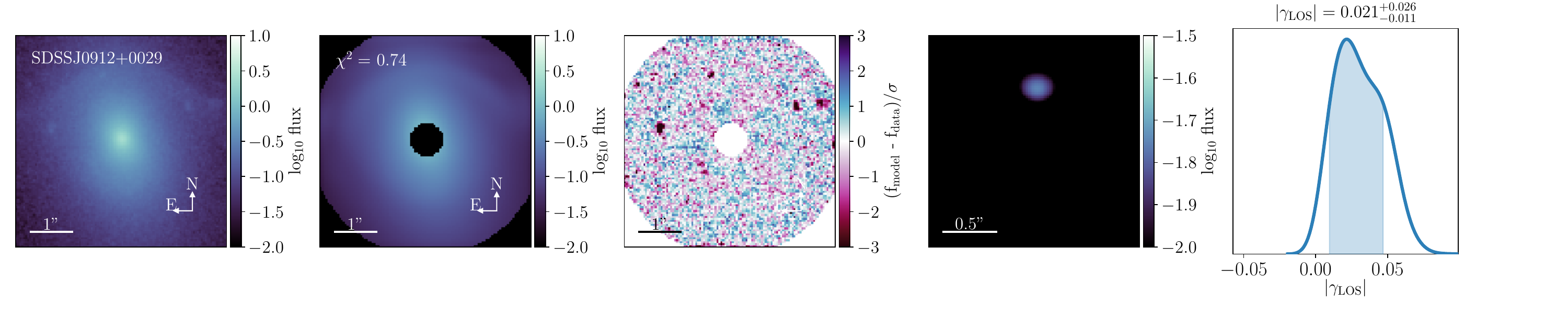}\\
    \includegraphics[width=\textwidth]{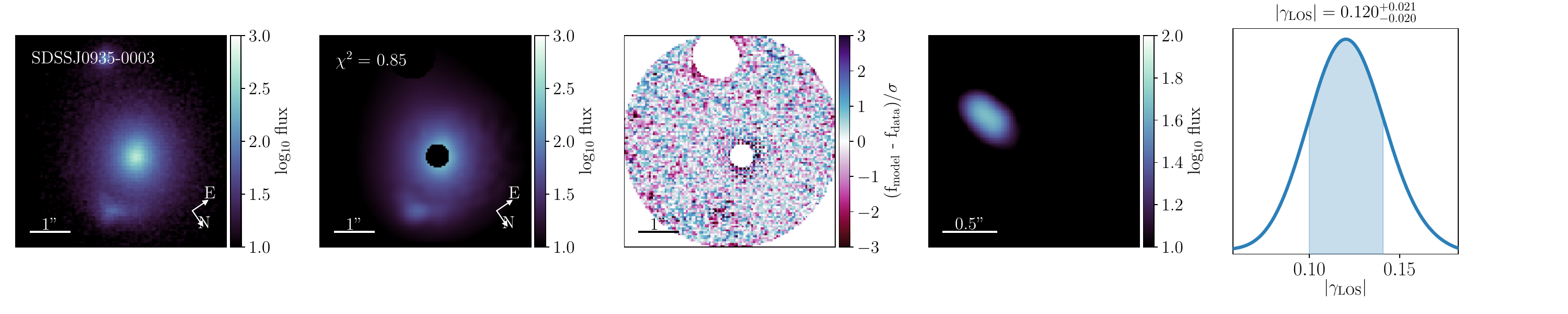}\\
    \includegraphics[width=\textwidth]{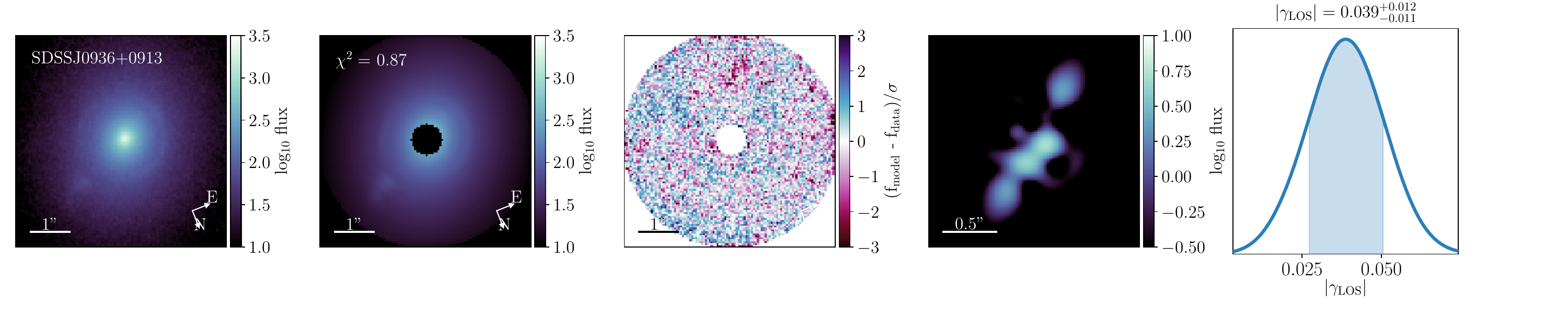}
	\caption{The first six lenses fit with the minimal model. From left to right, the panels show the single-band image data for each lens, our reconstruction of the image along with the reduced $\chi^2$ of the model, the residual difference between the image and the reconstruction, the reconstructed source and the one dimensional marginalised posterior distribution of the LOS shear magnitude, $|\gamma_{\rm LOS}|$. The shaded area is the $1 \sigma$ confidence interval.}
	\label{fig:min1}
\end{figure*}

\begin{figure*}
	\centering
	\includegraphics[width=\textwidth]{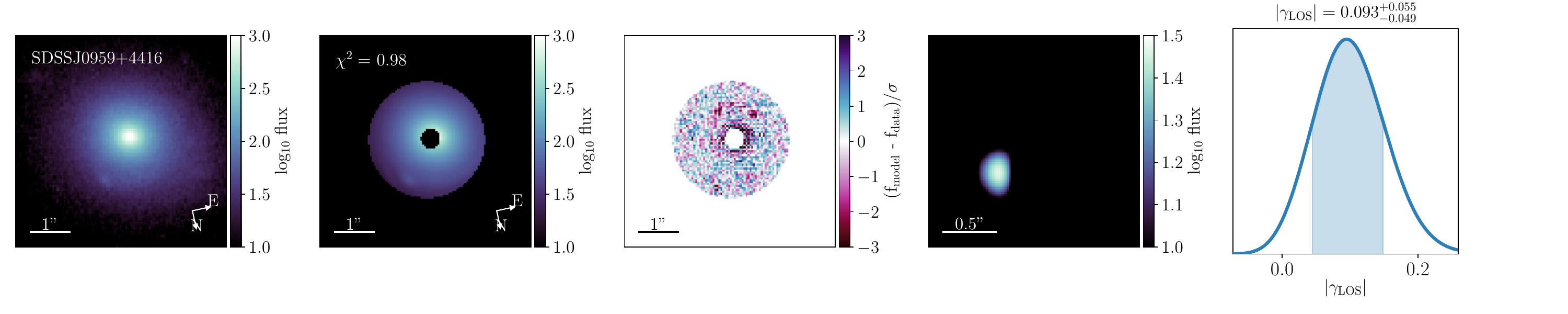}\\
	\includegraphics[width=\textwidth]{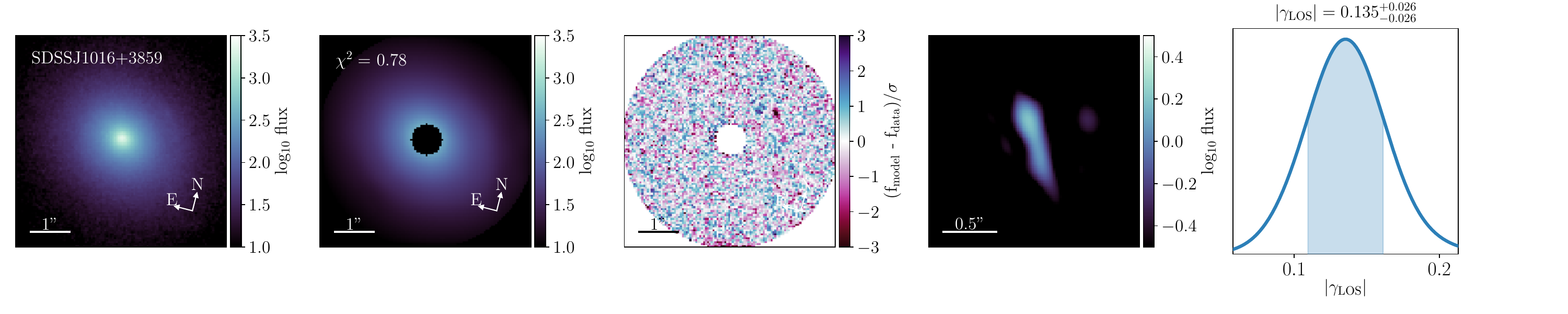}\\
	\includegraphics[width=\textwidth]{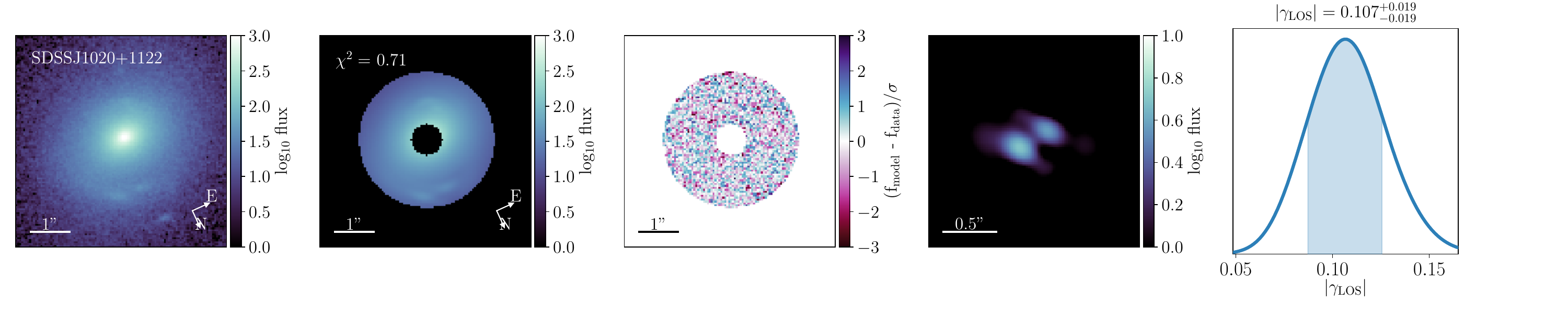}\\
	\includegraphics[width=\textwidth]{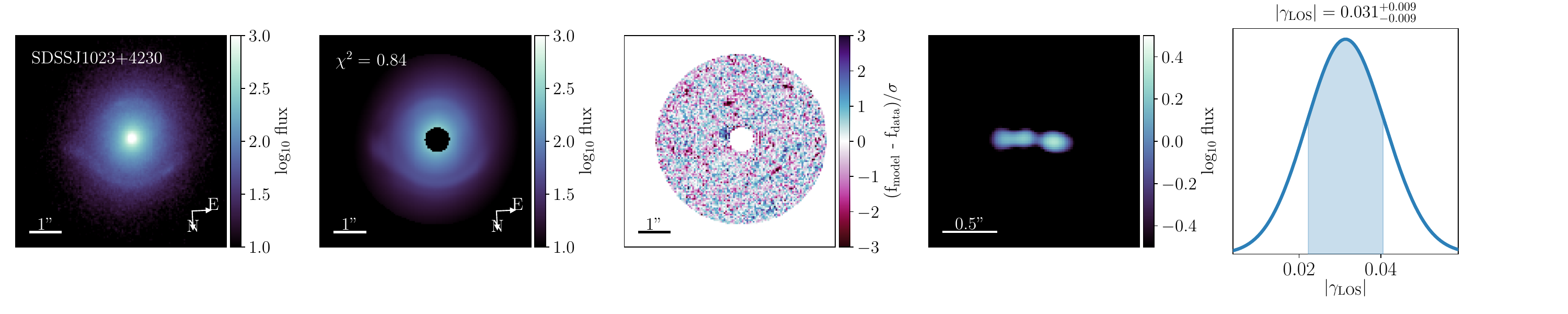}\\
	\includegraphics[width=\textwidth]{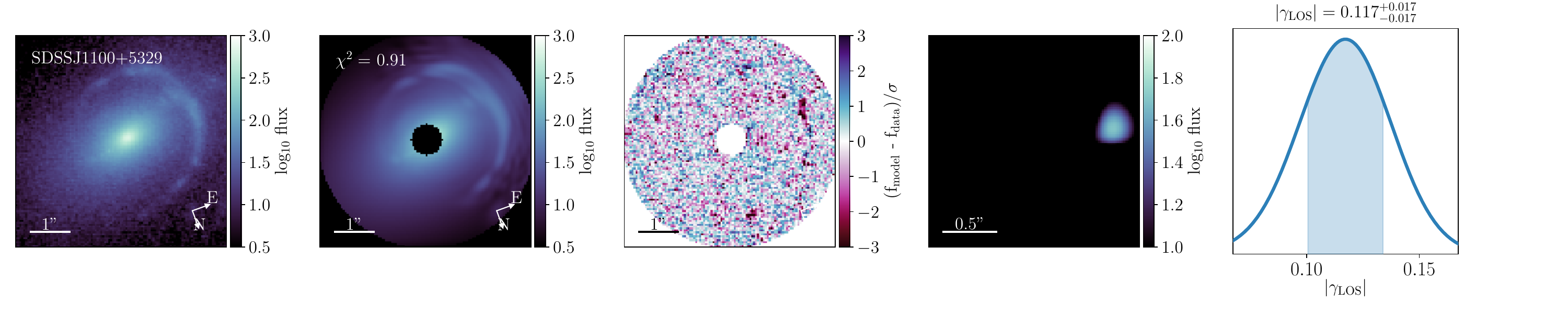}\\
	\includegraphics[width=\textwidth]{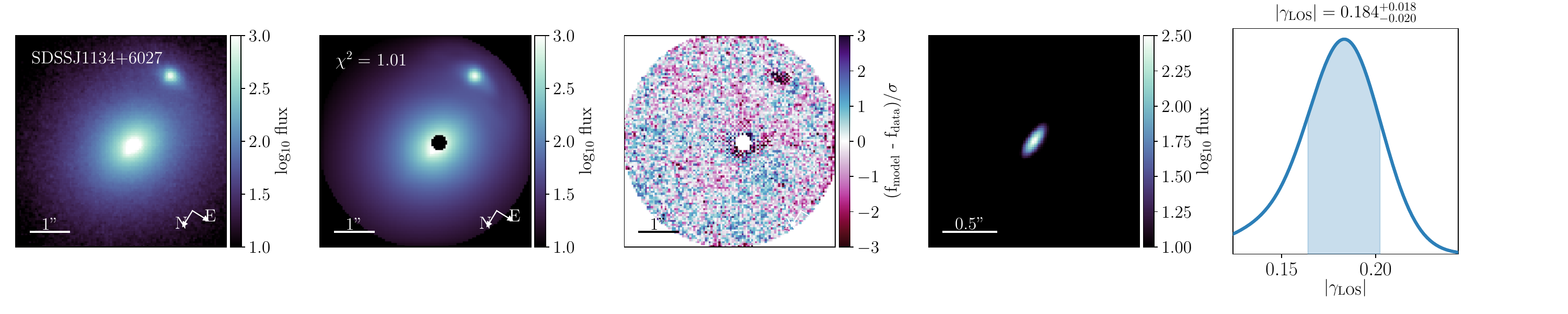}
	\caption{The next six lenses fit with the minimal model. The panels show the same information as in \autoref{fig:min1}.}
	\label{fig:min2}
\end{figure*}

\begin{figure*}
	\centering
	\includegraphics[width=\textwidth]{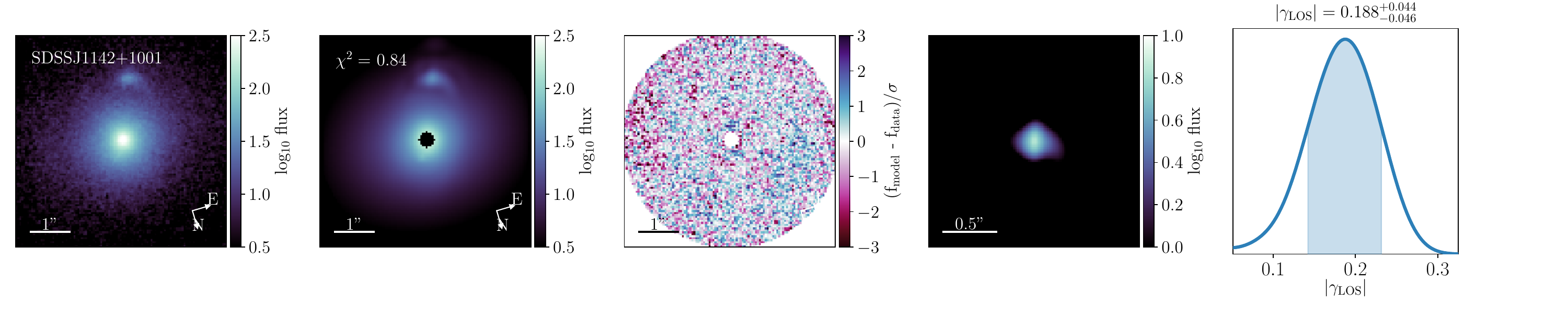}\\
	\includegraphics[width=\textwidth]{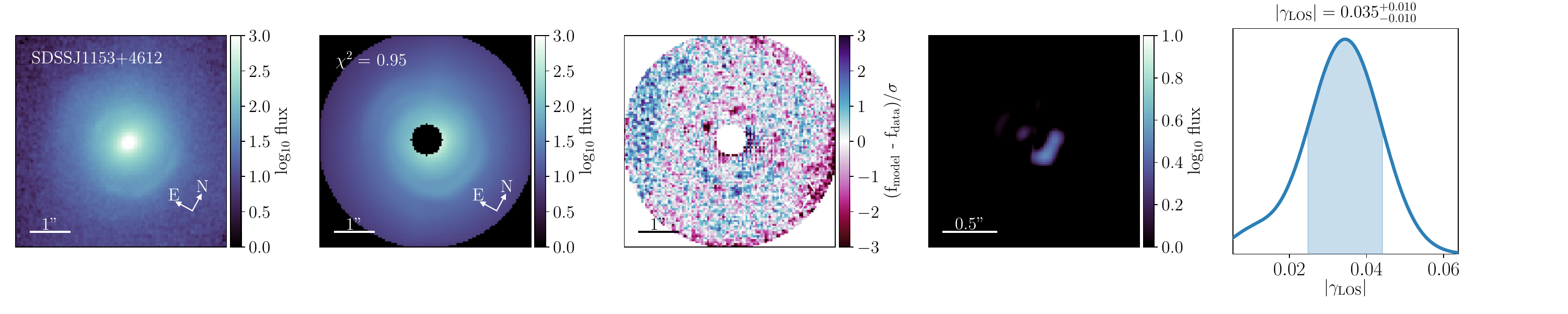}\\
	\includegraphics[width=\textwidth]{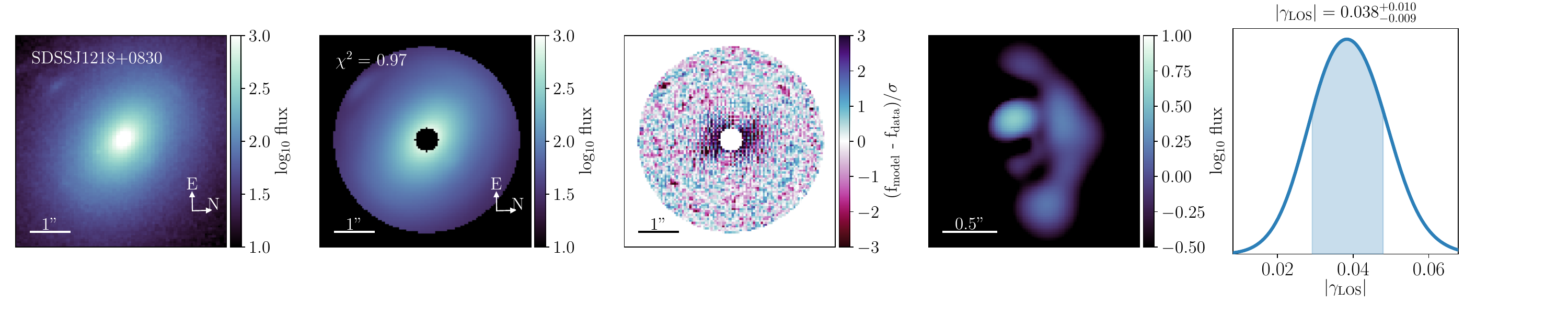}\\
	\includegraphics[width=\textwidth]{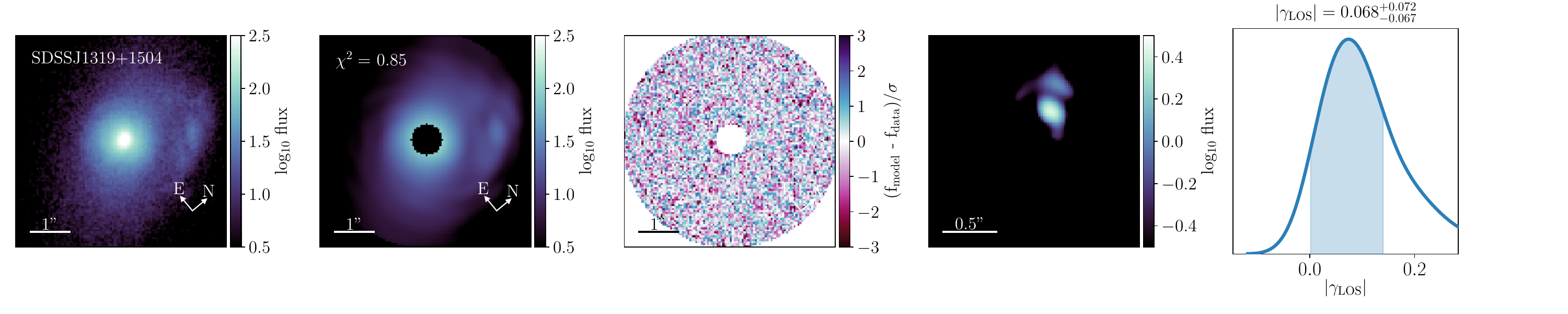}\\
	\includegraphics[width=\textwidth]{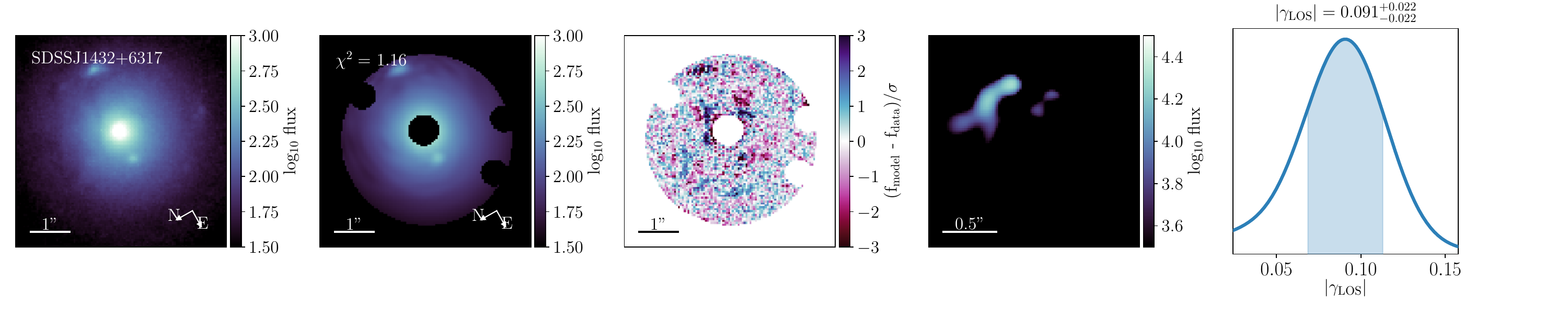} \\
    \includegraphics[width=\textwidth]{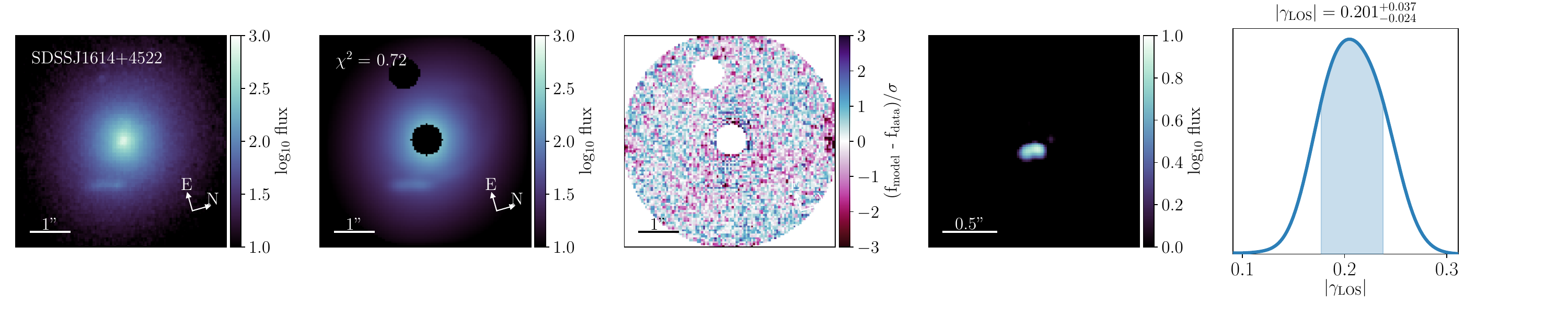}
	\caption{The next six lenses fit with the minimal model. The panels show the same information as in \autoref{fig:min1}.}
	\label{fig:min3}
\end{figure*}

\begin{figure*}
	\centering
	\includegraphics[width=\textwidth]{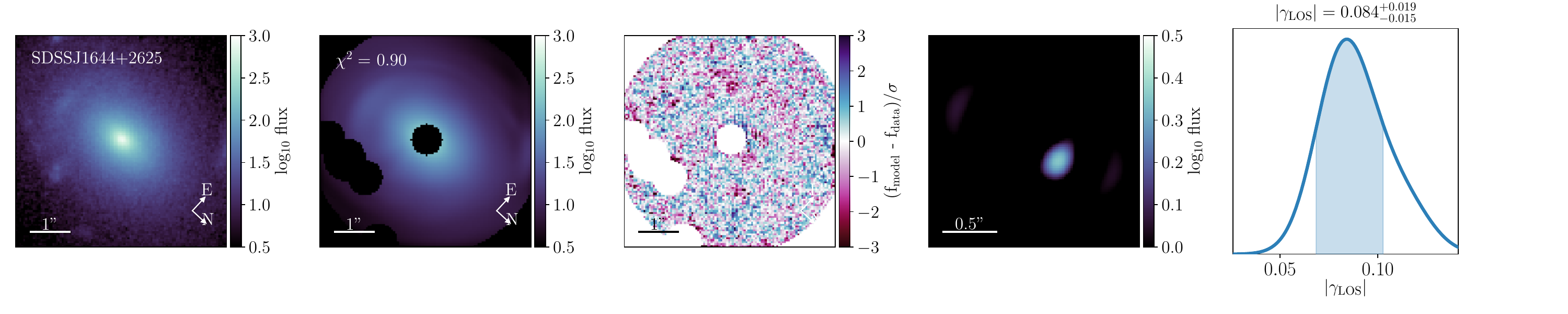}\\
	\includegraphics[width=\textwidth]{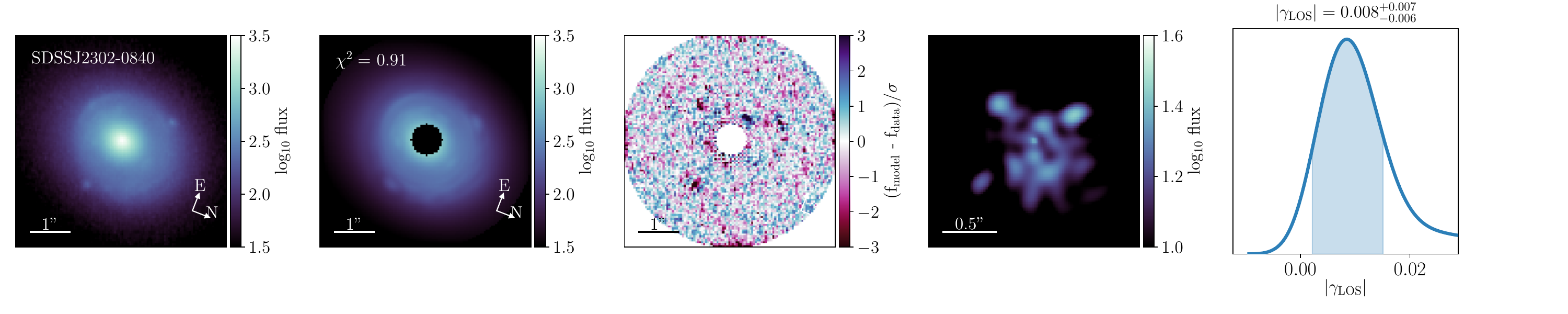}\\
	\includegraphics[width=\textwidth]{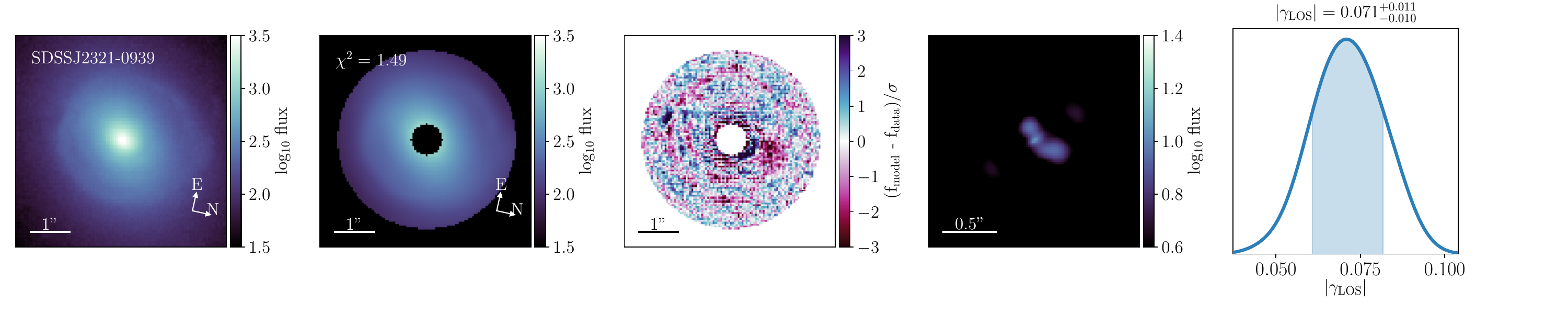}\\
	\includegraphics[width=\textwidth]{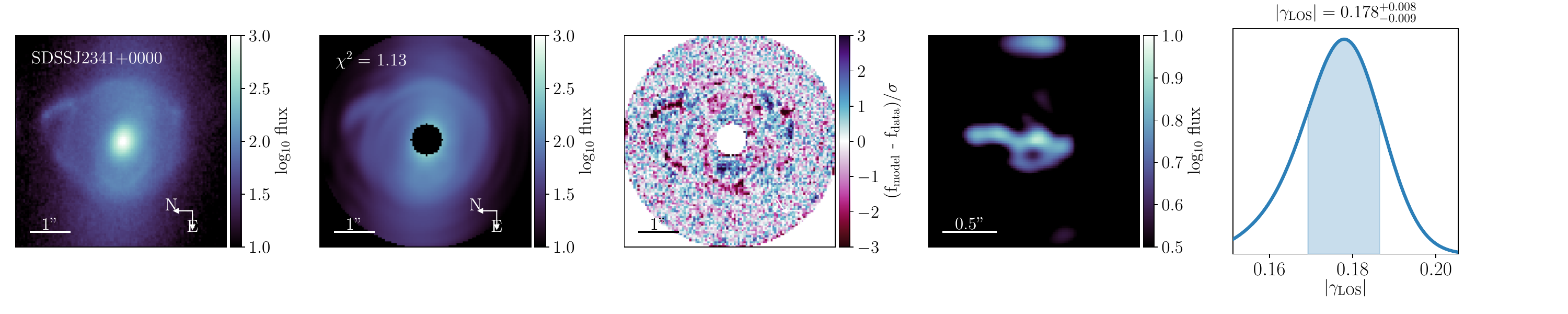}\\
	\caption{The final four lenses fit with the minimal model. The panels show the same information as in \autoref{fig:min1}.}
	\label{fig:min4}
\end{figure*}

We successfully modelled 22 lenses out of 27; meaning that 45 out of our initial sample of 50, or 90\% of selected strong lenses from the SLACS programme have been successfully modelled in a semi-automated fashion using the minimal LOS shear parameterisation. In each column of \autoref{fig:min1}--\autoref{fig:min4}, we show the image data for each lens, our model reconstruction, the normalised residuals between the data and model, the source reconstruction and finally the one-dimensional marginalised posterior distribution for $|\gamma_{\rm LOS}|$, computed from the posterior samples of its components, $\gamma_1^{\rm LOS}$, $\gamma_2^{\rm LOS}$. We find that the LOS shear is measured across this sample of lenses with a mean magnitude of $0.11\pm 0.024$. Including the 23 lenses studied in Paper I, the mean magnitude across the sample of 45 lenses is $0.085 \pm 0.019$.

\begin{figure*}
	\centering
    \includegraphics[width=\textwidth]{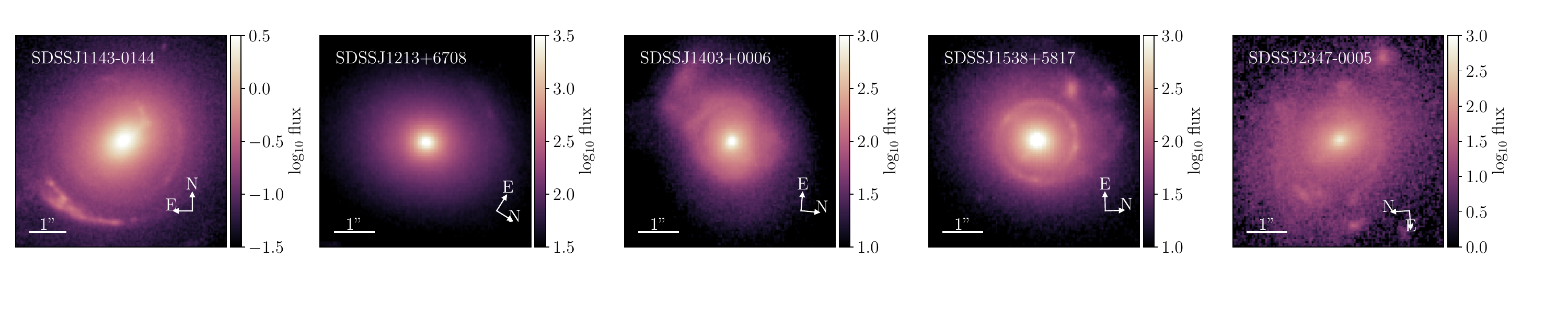}
\caption{The lenses in our sample for which we consider the modelling using the baseline EPL + minimal LOS shear model to be a failure.}
	\label{fig:failures}
\end{figure*}

The following lenses, shown in \autoref{fig:failures}, are excluded from our list of successes:
\begin{itemize}
    \item \textit{SDSSJ1143$-$0144}: this lens exhibits a complex morphology with two concentric arcs, the inner significantly fainter than the outer. We attempted to reconstruct the arcs individually, masking the other arc in turn, but in both cases, several parameters in our MCMC did not converge.
    \item \textit{SDSSJ1213+6708}: the relatively faint lensed source emission compared to the lens light led to poor image reconstruction and convergence in our MCMC chains.
    \item \textit{SDSSJ1403+0006}: the strong blending of the lensed source emission with the light of the main deflector in this case presented a particular challenge with this lens, leading to a lack of convergence in our MCMC chains.
    \item \textit{SDSSJ1538+5817}: again, blending between the lens light and lensed source emission led to a poor source reconstruction in this lens.
    \item \textit{SDSSJ2347$-$0005}: this system has a number of foreground or satellite features associated with the main deflector, some of which we were able to mask out, but the forward modelling did not produce a convincing source reconstruction.
\end{itemize} 

\begin{figure}
    \centering
    \includegraphics[width=0.49\textwidth]{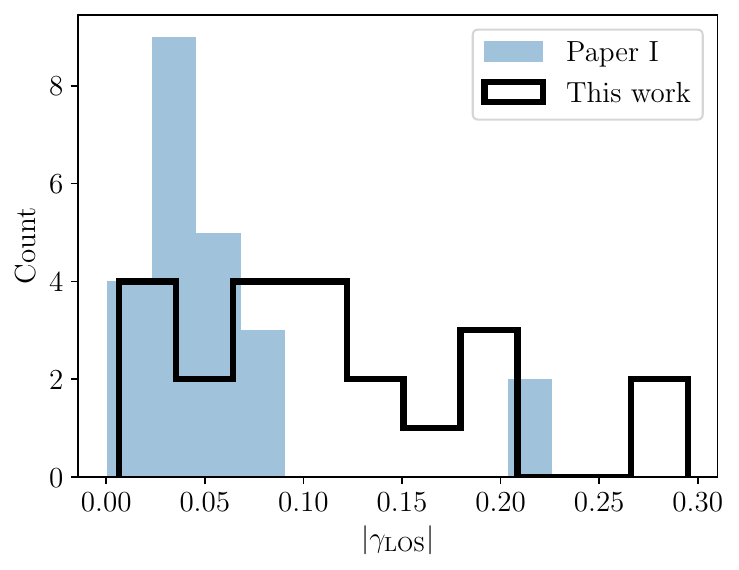}
    \caption{Histograms of the $|\gamma_{\rm LOS}|$ values measured from the lenses presented in Paper I (solid blue) and in this work (unfilled black).}
    \label{fig:histogram}
\end{figure}

\autoref{fig:histogram} shows the distribution of the $|\gamma_{\rm LOS}|$ shear measurements made in this work (unfilled black) and in Paper I (solid blue). A significant proportion of the lenses in this work yield larger shear magnitudes than those in Paper I; here, 12 out of 22 have $|\gamma_{\rm LOS}| > 0.1$, compared to only two out of 23 in Paper I.

\subsection{Large shear magnitudes}
We may therefore ask, as in Paper I, what is driving these large shear magnitudes? The hypothesis previously tested in that work and in the literature is that unmodelled angular features, specifically an octupole (e.g. \cite{Etherington:2023yyh}),
act to boost shear values. Whilst shear and octupolar distortions have distinct mathematical effects on a lens mass and images, imperfectly degenerate parameters may nonetheless compensate for missing complexity in a model in unexpected ways; see, for example, \cite{Etherington:2023yyh,Kochanek_2021,Hogg:2022ycw}. For the lenses studied in this work, we again investigated what happened when we fit them with the inclusion of an octupolar distortion in the lens mass model, specifically, the \texttt{EPL\_BOXYDISKY} profile available in \lenstronomy \citep{VandeVyvere2021}.

We find that, for these lenses, the octupole inclusion often results in pathological problems in the recovered lens models, such as unphysically shallow EPL slopes ($\gamma_{\rm EPL} < 1.4$), or a complete lack of convergence in lens model parameters. We indicate which lenses were successfully modelled with the octupole in the last column of \autoref{tab:lenses}; there are nine in total.

\begin{figure*}
    \centering
\includegraphics[width=\textwidth]{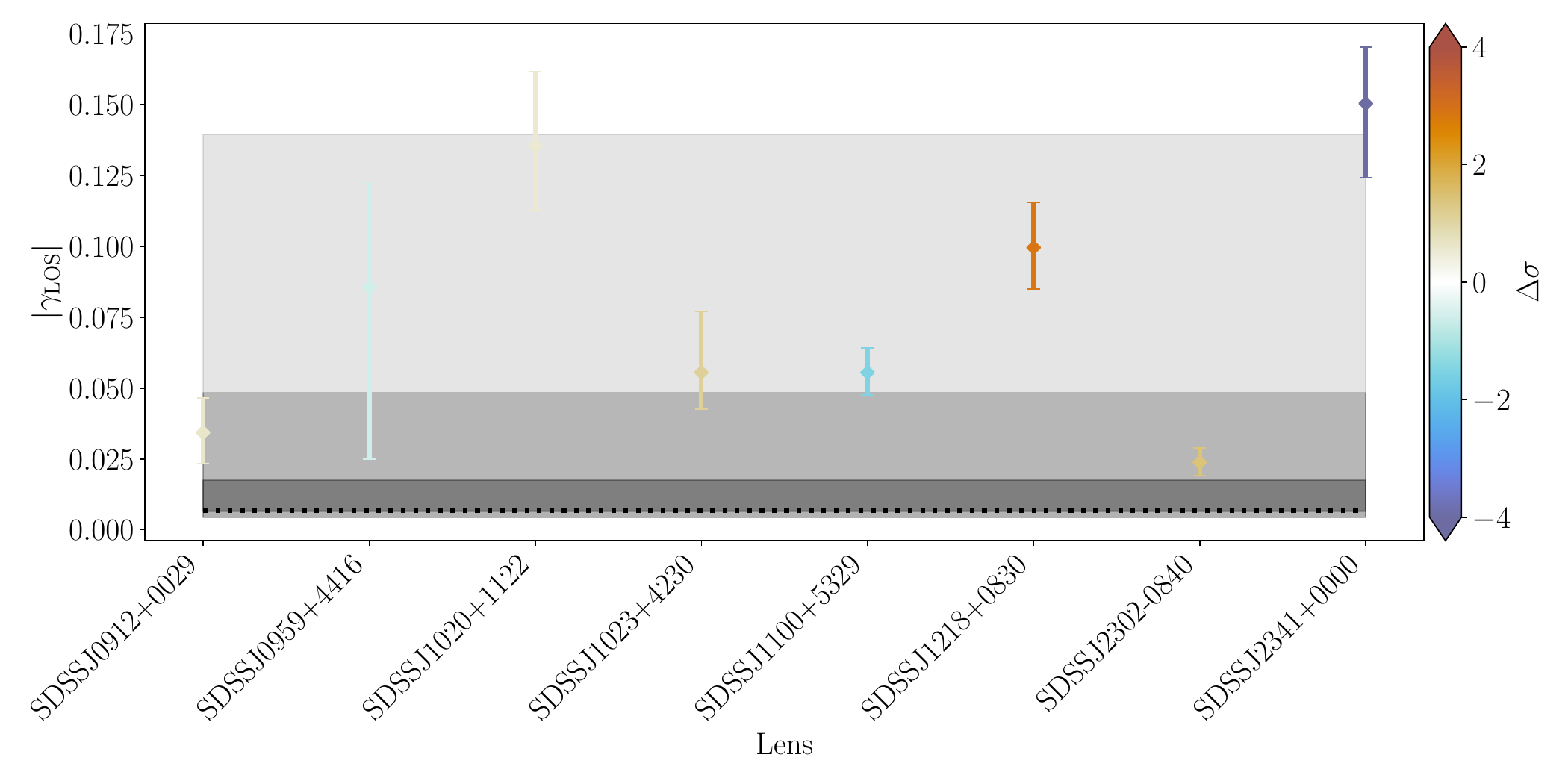}
    \caption{Each point in this figure shows the measured value of $|\gamma_{\rm LOS}|$ with the octupole included in the lens mass model. The black dotted line shows the overall median expected value of  $|\gamma_{\rm LOS}|$ from an $N$-body simulation, with the associated 1, 3 and $5 \sigma$ uncertainties around this median given by the grey shaded bands. The measured values are coloured by the change in the tension between the minimal model and the minimal + octupole model. The systems are ordered by increasing lens redshift on the $x$-axis. Note that, by plotting the overall median expectation and uncertainties of the shear magnitudes from simulations, we neglect the trend with redshift (cf. Fig. 6 in Paper I), but this trend is small compared to the shear values measured for these lenses.}
    \label{fig:tension}
\end{figure*}

As before, we find no systematic decrease in shear magnitude. In \autoref{fig:tension}, we show the change in tension of the shear measurement compared to the expectation from an $N$-body simulation\footnote{See Paper I and \cite{Johnson2025} for a full description of this method. The simulations used are the \texttt{RayGalGroupSims} \citep{2019MNRAS.483.2671B,Rasera:2021mvk}.}, for those eight lenses with successful octupole models,
\begin{equation}
    T_{ab} = \frac{|x_a - x_b|}{\sqrt{\rule{0pt}{2ex}\sigma_a^2 + \sigma_b^2}},
\end{equation}
where $x$ are the best-fit values and $\sigma$ the standard deviations of the two quantities being compared, for the lenses with successful octupole measurements, compared to the shear measurement in the minimal model. SDSS1213+6708 is excluded from this figure as it was not successfully modelled with the minimal model.

From this figure, we see that there is only a single system for which the tension with the expectation from the simulations decreases by more than $1.5\sigma$, the lens SDSSJ2341+0000, which exhibits a decrease in tension $\Delta \sigma = -7.0$. There is a small reduction in reduced $\chi^2$ compared to the model without the octupole, $\Delta \chi^2 = -0.02$,  whilst the octupole strength is inferred as $a_4 = -0.047^{+0.019}_{-0.015}$. The precision on the shear measurement is identical in each case, $\pm 0.009$, and the position angles of the shear are consistent at $1\sigma$. Nevertheless, the shear magnitude inferred in the presence of the octupole is $|\gamma_{\rm LOS}| = 0.133\pm 0.009$, still outside the $5\sigma$ uncertainty on the median expected $|\gamma_{\rm LOS}|$ shear from the $N$-body simulation. Therefore, as in Paper I, we have very little evidence in support of the hypothesis that an unmodelled octupole systematically acts to boost inferred shear magnitudes. Other types of angular complexity, such as isophotal twistyness, remain as avenues for future exploration.

\subsection{Shear, mass and light are randomly orientated}

\cite{Etherington:2023yyh} found preferential alignment between external shear and the lens mass in both simulations and data. We compare the position angles of the inferred LOS shear with that of the lens mass ellipticity, and the ellipticity of the two S\'{e}rsic profiles we use to model the lens light. To improve the statistical power for this test, we examine the lenses modelled in this work along with those of Paper I.

\begin{figure}
    \centering
    \includegraphics[width=0.5\textwidth]{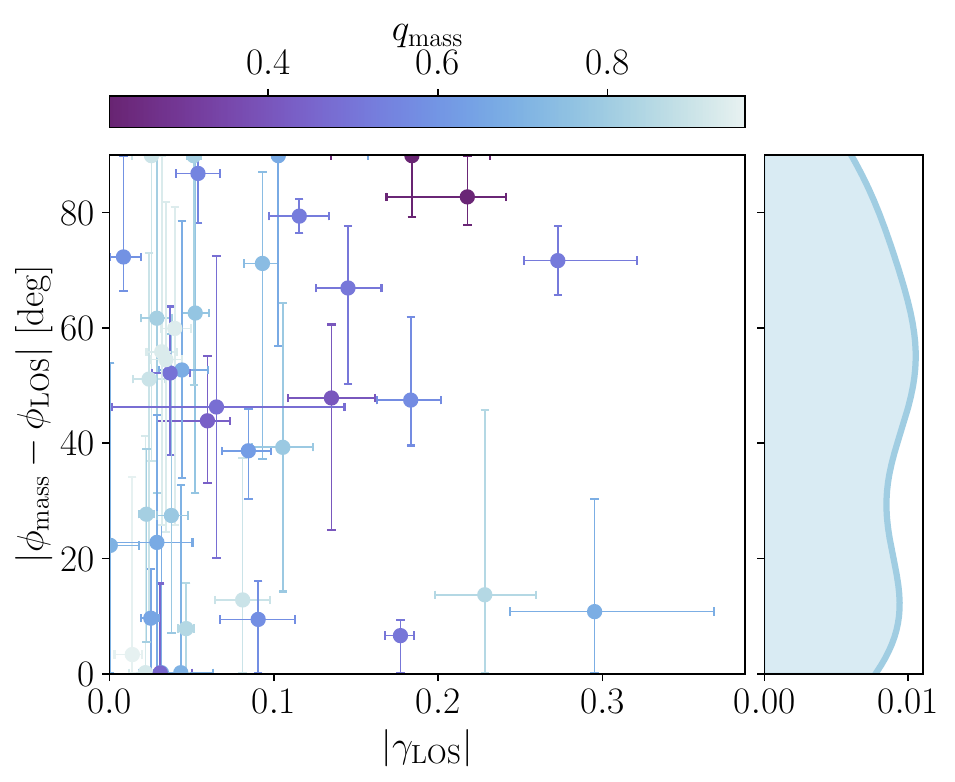}
    \caption{Relative orientation of the lens mass ellipticity and LOS shear for the combined Paper I and Paper II (this work) sample of lenses. The points are coloured by the axis ratio of the lens mass. The right-hand panel shows a KDE of the distribution of $|\phi_{\rm mass}- \phi_{\rm LOS}|$.}
    \label{fig:alignment_scatter}
\end{figure}

We plot the absolute difference between the position angle of the mass and LOS shear against the LOS shear magnitude, $|\phi_{\rm mass}- \phi_{\rm LOS}|$ in \autoref{fig:alignment_scatter}, with the points coloured by the axis ratio of the mass component. Significant alignment or anti-alignment of the two components would see the points clustered near $0 \degree$ or $90 \degree$; instead, the points are scattered uniformly across the range of angles. In the right-hand panel of this figure, we show a kernel density estimation (KDE) for the distribution of $|\phi_{\rm mass}- \phi_{\rm LOS}|$. A Kolmogorov--Smirnov test confirms that this distribution does not depart significantly from uniformity, as expected for random component alignments, with a KS statistic of 0.114 and a p-value of 0.56. The same finding holds for the relative orientations of the LOS shear with the two individual lens light components; the respective KS statistics and p-values for these distributions are 0.109 (0.62) and 0.093 (0.79).

We note that for the lenses with the very largest shear magnitudes ($|\gamma_{\rm LOS}|\gtrsim 0.2$), there appears to be some indication from \autoref{fig:alignment_scatter} that the LOS shear is aligned/anti-aligned with the lens mass. We also see that the most elliptical lenses (smallest $q_{\rm mass}$, darker coloured points) tend towards anti-alignment of the two components ($|\phi_{\rm mass}- \phi_{\rm LOS}|>45 \degree$). The two lenses with the combination of largest shears, highest anti-alignment, and smallest axis ratios are SDSSJ1531$-$0105 and SDSSJ1142+1001. The latter was not successfully modelled with the inclusion of an octupole in the lens mass, but the former was (see Paper I), which resulted in a significant reduction in `tension' with the expected shear magnitude from simulations for that lens. An anti-aligned shear and ellipticity are morphologically similar to a pure octupole. However, we again stress that the inclusion of the octupole does not \textit{systematically} reduce the LOS shear magnitude across this sample of lenses.

\subsection{What do large and small shear lenses have in common?}
If the question of large shears cannot be solved by the introduction of an octupole, it is apposite to attempt to quantify what features the lenses with large shears have in common with each other that may distinguish them from lenses with small shears, beyond the mass distribution of the main deflector, and which are not features derived from the lens modelling. To do this, we split our 45 modelled lenses into two subsets based on the median inferred shear value, $|\gamma_{\rm LOS}| = 0.052$, leading to 22 lenses with `small' shear (less than the median) and 23 lenses with `large' shear (equal to or larger than the median).

\subsubsection{Redshift and sky location}
A priori, the two subsets of lenses are identical in their selection, all having been detected as candidate lenses via the high-redshift emission lines in the sources seen in Sloan Digital Sky Survey (SDSS) data. This means that spectroscopic redshifts are available for all lenses and sources (listed in \autoref{tab:lenses}). Before considering any features of the photometry or lens morphology, we can check whether the large shear lenses have significantly higher redshifts, as a longer line of sight would be expected to possess more structure, and therefore potentially induce a larger shear on a strong lens image.

To do this, we perform the two-sample Kolmogorov--Smirnov (KS) test \citep{Sprent1998}, which tests the null hypothesis that two samples are drawn from the same underlying distribution. In our case, the two samples are the lens redshifts for the large shear lenses and the small shear lenses; and then the source redshifts for the large shear and small shear lenses. In both cases, the null hypothesis is sustained, with p-values of 0.55 and 0.71 respectively; in other words, there is no significant evidence of lenses with large shear values having a different distribution in redshift to those with small shear values. We note that all of the SLACS lenses and sources lie in a relatively restricted range of redshifts, so this question could be revisited with a higher-redshift population, such as the ``COWLS'' catalogue of strong lenses \citep{Nightingale:2025mlk, Mahler:2025ajl, Hogg:2025plt} discovered in COSMOS-Web \citep{Casey:2022amu}.

The SLACS lenses in our sample are very well distributed across the sky, with an average angular separation of $82.5\degree$. The closest pair of lenses, SDSSJ2343$-$0030 and SDSSJ2341+0000, are $0.71\degree$ apart. There is no significant clustering of large shear and small shear lenses in a given area of the sky.

\subsubsection{Filters and PSF}
After detection in SDSS, the lens candidates were followed up with HST. The photometry available to us in this work was a single HST band per lens, either F555W or F606W. As with the redshifts, we can check to see whether large shears are preferentially inferred when using one or other of the filters.

Across the 45 lenses, we have imaging data for 35 in F606W and 10 in F555W; in other words, $77.8\%$ of the lenses were modelled using F606W, meaning that, if filter choice is irrelevant, we should expect $77.8\%$ of each of the large and small shear subsets to be F606W images. However, what we actually see is that nearly all of the large shear lenses (22/23, or $95.7\%$) are imaged in F606W. Using a binomial test to examine the significance of this, we find a p-value of 0.042, implying the correlation between large shear and filter choice is actually significant. It is clear that this is not a \textit{physical} correlation, but likely a result of how well lens features are resolved, and particularly how well lensed source emission and lens light are separable, across the two filters.

\begin{figure}
    \centering
\includegraphics[width=0.48\textwidth]{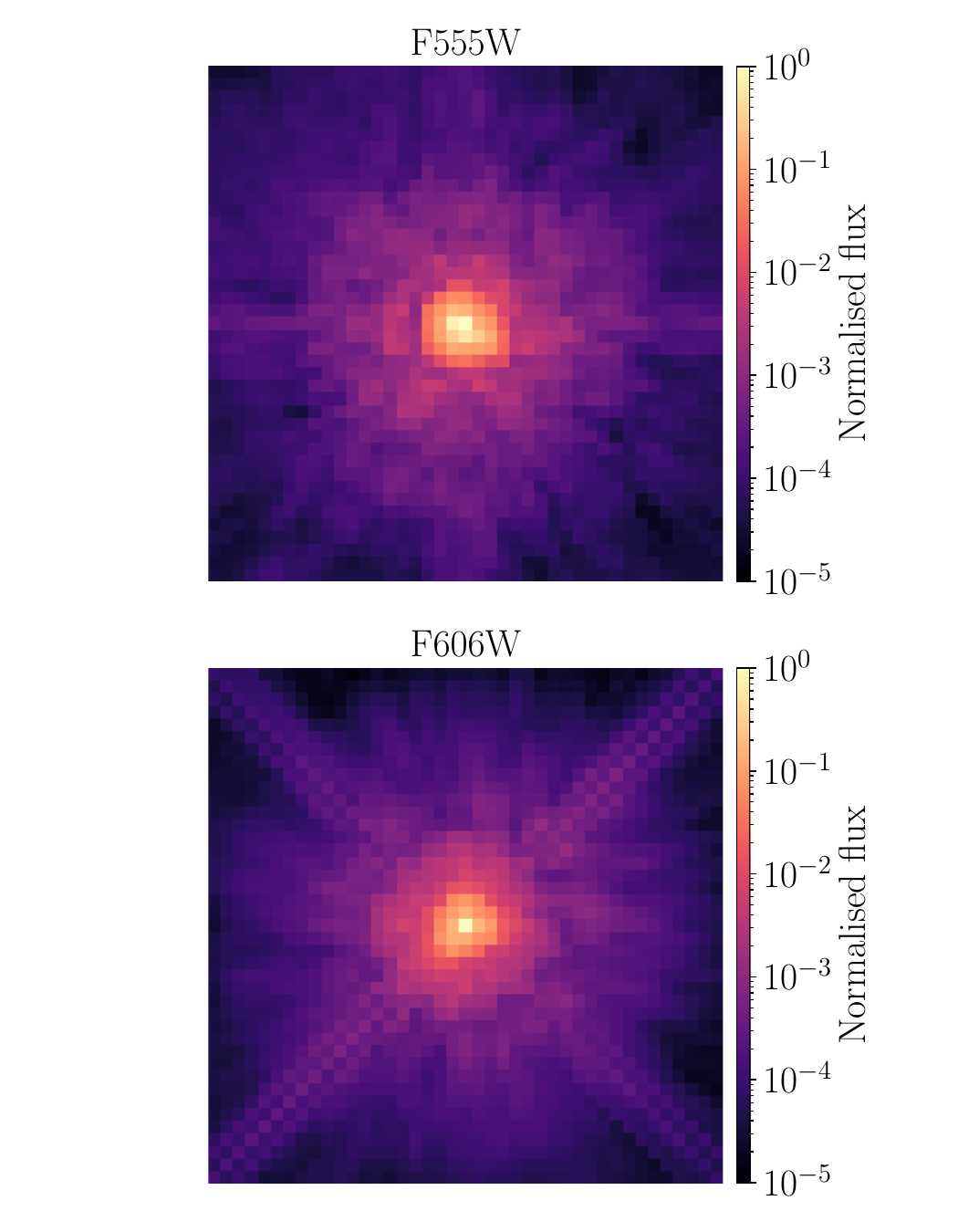}
    \caption{The point spread functions (PSFs) associated with the two HST filters in which the lenses studied in this work were observed. The colour bar indicates the flux in a given pixel of the PSF kernel, normalised so the maximum flux is one.}
    \label{fig:psfs}
\end{figure}

The other factor at play related to the filters of our observations is the PSF. The PSF for the F606W observations is somewhat sharper than F555W, but has an elongated tail, see \autoref{fig:psfs}. To test the effect of the PSF on shear recovery, we simulated 100 images with the F555W PSF, and 100 images with the F606W PSF. We used a simplistic lens model consisting of a Singular Isothermal Ellipsoid plus external shear, with an elliptical S\'ersic profile for the lens and source light. For each of the 200 images, the model parameters were drawn at random. We then fit these images with a PSO+MCMC, and examined the difference between the input and recovered shear values.

\begin{figure}
    \centering
    \includegraphics[width=0.48\textwidth]{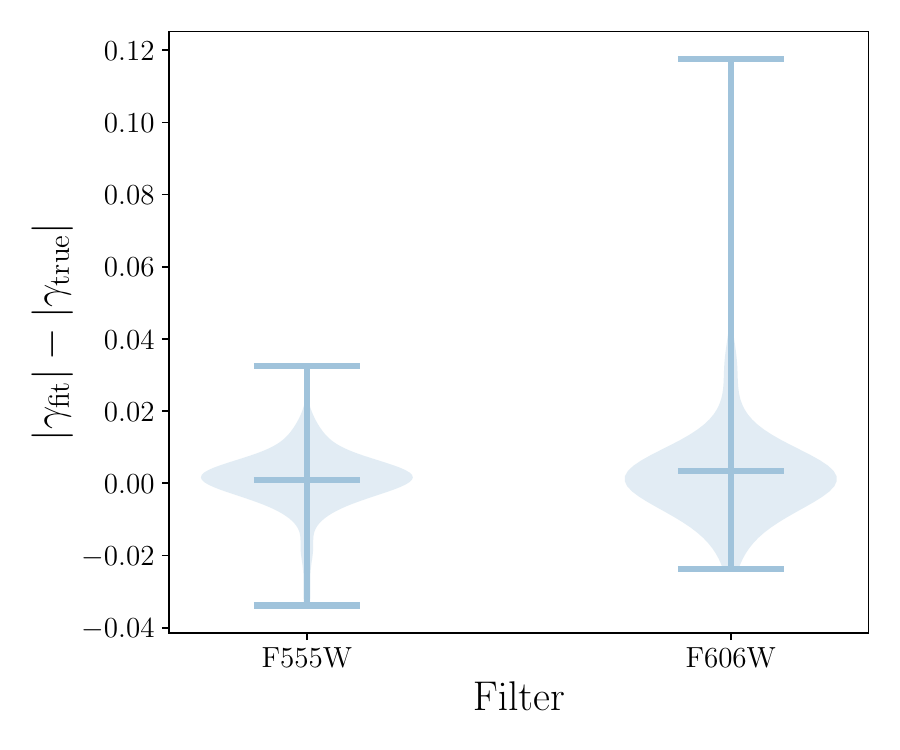}
    \caption{Violin plots showing the distribution of the difference between the value of $|\gamma|$ recovered after fitting the mock lenses and the value input to create the mocks, split by filter.}
    \label{fig:violins}
\end{figure}

\autoref{fig:violins} shows violin plots for the distribution of the difference between the recovered and input values of the shear magnitude in our simple mocks, or the `shear residual', using the two different PSFs associated with the two filters. The horizontal bars show the minimum, mean, and maximum values for each distribution. From this figure, we can see that the shear residual distribution for F606W is positively skewed compared to the F555W distribution, with a much higher maximum shear residual. However, again using the KS test, we cannot reject the null hypothesis that the two samples follow the same underlying distribution ($p=0.58$).

\cite{Shajib2021} also tested the effects of different PSF estimation methods on their inference of the logarithmic slope of the lens mass profiles (these lenses are the same as studied in our Paper I) and found there was negligible effect due to PSF choice.

\subsubsection{Flux and signal-to-noise ratio}
Lastly, we can test photometric features of the images in the large shear and small shear groups. Specifically, we use the tailored masks produced for each lens (shown in the second column of \autoref{fig:min1}--\autoref{fig:min4}) to define the region of interest in each image where we will examine the flux and signal-to-noise ratio (SNR), as these are the regions which contain primarily lensed source emission.

Lenses in the large shear group tend to have slightly larger masks (on average 3522 pixels masked versus 3342 in the small shear group), meaning that they have fewer pixels which are used in the likelihood evaluation, which leads to a greater uncertainty on the best-fit shear derived from the posterior. The large shear lenses have a mean flux in their arcs which is nearly twice as high as that of the small shear lenses, 43.8 e$^-$/s versus 22.2 e$^-$/s. The p-value of the KS test in this case is 0.069, approaching statistical significance. Examining the average SNR across the arcs in our two subsets of lenses, we find no difference between the large shear and low shear lenses. Accordingly, the large shear subset has a larger background RMS, which counteracts the flux to produce a similar SNR distribution to the small shear subset. We therefore cannot conclude that SNR strongly influences the recovered shear magnitude.

\section{Conclusions} \label{sec:conclusions}
In this work, a companion to our Paper I \citep{Hogg:2025wac}, we modelled 27 strong gravitational lenses from the SLACS catalogue, achieving successful lens models for 22 of these systems. Added to the sample of 23 modelled in Paper I, this makes a total of 45 strong gravitational lenses modelled with the goal of measuring the LOS shear in these systems.

We found that the mean LOS shear magnitude across the lenses studied in this work is  $|\gamma_{\rm LOS}| =  0.11\pm 0.024$; when the lenses from Paper I are included, this decreases to $|\gamma_{\rm LOS}| =  0.085 \pm 0.019$. Both of these figures are significantly larger than the LOS shear magnitude computed by ray-tracing through $N$-body simulations, a finding which correlates with other studies of shear in strong lenses \citep{Etherington:2023yyh}. In contrast to \cite{Etherington:2023yyh}, however, we do not find a preferential alignment between the measured shear and the measured lens mass or light ellipticity.

In Paper I we tested whether the inclusion of an octupolar distortion in the lens mass could act to reduce the inferred shear magnitude; we repeated this test for the lenses studied in this work, finding in both cases that the octupole does not result in a systematic reduction in shear magnitude. 

Consequently, we tested other features of the lenses themselves, distinct from questions related to lens modelling. Splitting our sample of 45 lenses into two subsets, those with shears larger than the median, and those with shears smaller than the median, we investigated whether these two populations showed evidence of being drawn from different redshift or sky location distributions, finding no such evidence.

We further tested the correlation between the filter the lenses were observed with and the shear, finding a statistically significant preference for lenses observed in the F606W band to belong to the large shear subset. We investigated whether this could be due to the different PSF shape between the two filters by simulating and fitting simple mock images with the two PSFs, but we found no statistically significant difference between the two. 

Lastly, we checked whether the two subsets of lenses showed commonalities in flux or SNR. We found that lenses with large shears tend to have a substantially larger mean flux (counting only the unmasked regions) but also a larger background RMS, the consequence of which being that the distributions of SNR across the two subsets is not significantly different.

We therefore conclude that there is no obvious feature in the photometry of a given lens that leads to an unexpectedly large shear being inferred; in other words, there is no clear systematic bias introduced by the observations themselves. The between-filter difference we observe likely arises from the corresponding image quality, with certain features being visible in one band rather than another, increasing or reducing the importance of imperfect degeneracies between the shear and properties of the lens and source. This may be further investigated in samples of lenses with high-quality multi-wavelength data, such as COWLS~\citep{Nightingale:2025mlk}.

The answer to the conundrum of large shears must lie in the lens and source model themselves, but where is not obvious, given the fact that current models reconstruct strong lensing images down to the noise level. One of the biggest challenges strong lens modelling has ever faced lies ahead: what level of model complexity is truly justified by the data? This question may be answered in the near future thanks to continued advancements in lens modelling techniques
\citep{PerreaultLevasseur:2017ltk, Shajib:2019crn, Wagner-Carena:2020yun, Galan:2020mnn, Vernardos2022,Legin:2025abc}, hydrodynamical simulations which provide a testbed for these methods \citep{Mukherjee:2018cdf, Schaye:2025xuv}, and the wealth of new strong lenses which are being discovered by the \textit{Euclid} mission \citep{Euclid:2024jyk, Euclid:2025rfv}. Large samples of strong lenses and accurate shear measurements will in future enable precision cosmology, in cross-correlation with cosmic shear and galaxy clustering \citep{Fleury:2026pco}.

\section*{Acknowledgments}
We thank Pierre Fleury and Giacomo Queirolo for their comments on the manuscript and for valuable discussions surrounding this work. We are grateful to the anonymous referee of \cite{Hogg:2025wac} whose comments led to a substantial improvement of our results in that work and those presented in this manuscript. 

NBH is supported by the research environment and infrastructure of the Handley Lab at the University of Cambridge. DJ acknowledges support by the First Rand Foundation, South Africa, and the Centre National de la Recherche Scientifique of France.

\section*{Data and software availability}
The imaging data for the SLACS lenses studied in this work are publicly available. The lens modelling carried out in this work used the open-source \texttt{dolphin} package \citep{Shajib:2025bho}, with \lenstronomy \citep{Birrer:2018xgm, Birrer2021} as the modelling engine. The inputs and outputs of our modelling, including lens image cutouts, PSFs and MCMC chains, can be found at the following Zenodo repository: \url{https://zenodo.org/records/17816315}. A Jupyter notebook which reproduces the figures and tables in this work can be found at the following Github repository: \url{https://github.com/nataliehogg/los_in_slacs}.

The following colour-map packages were used in this work:  \texttt{cubehelix} \citep{Green2011},  \texttt{cmocean} \citep{Thyng2016} and \texttt{cmasher} \citep{vanderVelden2020}. NBH acknowledges the use of the AI coding agent Claude Code (Sonnet 4.5) for software development tasks. All AI-generated outputs were validated by NBH to ensure their accuracy.

\bibliographystyle{apsrev4-1}

\bibliography{los_slacs_ii_oja_accepted}

\appendix\label{appendix:config}
\renewcommand{\arraystretch}{1.8}

\begin{table*}[!ht]
    \centering
\begin{tabular}{llccccc}
\hline
\hline
Name & Filter & $n_{\rm max}$ & $z_{\rm lens}$ & $z_{\rm source}$ & $|\gamma_{\rm LOS}|$ & Octupole model? \\
\hline
SDSSJ0008$-$0004 & F606W & 6 & 0.440 & 1.192 & $0.295^{+0.073}_{-0.052}$ & \faClose \\
SDSSJ0044+0113 & F606W & 6 & 0.120 & 0.197 & $0.273^{+0.048}_{-0.020}$ & \faClose \\
SDSSJ0819+4534 & F606W & 16 & 0.194 & 0.446 & $0.145^{+0.020}_{-0.019}$ & \faClose \\
SDSSJ0912+0029 & F555W & 6 & 0.164 & 0.324 & $0.026^{+0.026}_{-0.012}$ & \faCheck \\
SDSSJ0935$-$0003 & F606W & 16 & 0.347 & 0.467 & $0.115^{+0.024}_{-0.025}$ & \faClose \\
SDSSJ0936+0913 & F606W & 6 & 0.190 & 0.588 & $0.037^{+0.012}_{-0.011}$ & \faClose \\
SDSSJ0959+4416 & F606W & 6 & 0.237 & 0.531 & $0.103^{+0.054}_{-0.050}$ & \faCheck \\
SDSSJ1016+3859 & F606W & 6 & 0.168 & 0.439 & $0.135^{+0.027}_{-0.027}$ & \faClose \\
SDSSJ1020+1122 & F606W & 6 & 0.282 & 0.553 & $0.105^{+0.019}_{-0.019}$ & \faCheck \\
SDSSJ1023+4230 & F606W & 6 & 0.191 & 0.696 & $0.032^{+0.009}_{-0.009}$ & \faCheck \\
SDSSJ1100+5329 & F606W & 16 & 0.317 & 0.858 & $0.115^{+0.018}_{-0.018}$ & \faCheck \\
SDSSJ1134+6027 & F606W & 6 & 0.153 & 0.474 & $0.183^{+0.018}_{-0.021}$ & \faClose \\
SDSSJ1142+1001 & F606W & 6 & 0.222 & 0.504 & $0.184^{+0.048}_{-0.049}$ & \faClose \\
SDSSJ1153+4612 & F606W & 6 & 0.180 & 0.875 & $0.034^{+0.010}_{-0.009}$ & \faClose \\
SDSSJ1218+0830 & F606W & 6 & 0.135 & 0.717 & $0.038^{+0.010}_{-0.009}$ & \faCheck \\
SDSSJ1319+1504 & F606W & 16 & 0.154 & 0.606 & $0.065^{+0.078}_{-0.064}$ & \faClose \\
SDSSJ1432+6317 & F606W & 16 & 0.123 & 0.664 & $0.090^{+0.022}_{-0.023}$ & \faClose \\
SDSSJ1614+4522 & F606W & 6 & 0.178 & 0.811 & $0.199^{+0.035}_{-0.024}$ & \faClose \\
SDSSJ1644+2625 & F606W & 6 & 0.137 & 0.610 & $0.084^{+0.019}_{-0.015}$ & \faClose \\
SDSSJ2302$-$0840 & F606W & 6 & 0.090 & 0.222 & $0.006^{+0.004}_{-0.003}$ & \faCheck \\
SDSSJ2321$-$0939 & F606W & 6 & 0.082 & 0.532 & $0.077^{+0.009}_{-0.012}$ & \faClose \\
SDSSJ2341+0000 & F606W & 16 & 0.186 & 0.807 & $0.177^{+0.008}_{-0.009}$ & \faCheck \\
\hline
SDSSJ1143$-$0144 & F555W & 16 & 0.106 & 0.401 & Unconstrained & \faClose \\
SDSSJ1213+6708 & F606W & 6 & 0.123 & 0.640 & Unconstrained & \faCheck \\
SDSSJ1403+0006 & F606W & 16 & 0.189 & 0.473 & Unconstrained & \faClose \\
SDSSJ1538+5817 & F606W & 16 & 0.143 & 0.531 & $0.042^{+0.016}_{-0.015}$ & \faClose \\
SDSSJ2347$-$0005 & F606W & 16 & 0.417 & 0.715 & $0.146^{+0.011}_{-0.011}$ & \faClose \\
\hline
\end{tabular}
\caption{The SLACS strong lenses studied in this work. All redshifts are taken from \cite{Bolton2008}. The $|\gamma_{\rm LOS}|$ values listed are those inferred in the minimal LOS shear model. The last column indicates if the lens in question was successfully modelled with the inclusion of an octupole in the lens mass.}
\label{tab:lenses}
\end{table*}

\begin{table*}
	\centering
\begin{tabular}{llll}
\hline
\hline
Component    & Parameter        &  Prior \\
\hline
	         & $\gamma^{\rm EPL}$         & $[1.3, 2.8]$ \\
          & $e_1$            & $[-0.5, 0.5]$\\
Lens mass & $e_2$            & $[-0.5, 0.5]$\\
	         & $x$              &  $[-0.5'', 0.5'']$\\
	         & $y$              & $[-0.5'', 0.5'']$\\
\hline
             & $R_{\rm{eff}}$   & $[0.1, 5.0]$\\
	         & $e_1$            & $[-0.5, 0.5]$\\
Lens light   & $e_2$            & $[-0.5, 0.5]$\\
	         & $x$              &  $[-0.5'', 0.5'']$\\
             & $y$              & $[-0.5'', 0.5'']$\\
\hline
	         & $\beta$          & $\log_{10} [0.02'', 0.2'']$\\
	         & $R_{\rm{eff}}$   & $[0.04'', 0.5'']$\\
	         & $n_{\rm{S}}$     & $[0.5, 8.0]$\\
Source light & $e_1$            & $[-0.5, 0.5]$\\
	         & $e_2$            & $[-0.5, 0.5]$\\
	         & $x$              & $[-0.2'', 0.2'']$\\
             & $y$              & $[-0.2'', 0.2'']$\\
\hline
	         & $\gamma_1^{\rm od}$ & $[-0.2, 0.2]$\\
	         & $\gamma_2^{\rm od}$ & $[-0.2, 0.2]$\\
Shear        & $\gamma_1\h{LOS}$   & $[-0.5, 0.5]$\\
             & $\gamma_2\h{LOS}$   & $[-0.5, 0.5]$\\
	         & $\omega\e{LOS}$     & $[-0.2, 0.2]$\\
\hline
                & $a_4$   & $[-0.1, 0.1]$\\
 Octupole       & $\phi_4$   & $[-\pi, \pi]$\\
	              & $x$      &  $[-0.5'', 0.5'']$\\
	              & $y$        & $[-0.5'', 0.5'']$\\

\hline
\end{tabular}
	\caption{Priors on the model parameters. The prior distributions are uniform between the stated limits.}
\label{tab:priors}
\end{table*}

\end{document}

%% file: packages.tex
\usepackage{graphicx}
\usepackage{amsmath}
\usepackage{gensymb}
\usepackage{xspace}
\usepackage[dvipsnames]{xcolor}
\usepackage{fontawesome}
\usepackage{bm}

\usepackage{hyperref}
\hypersetup{colorlinks=true, allcolors=Blue}

\newcommand{\lenstronomy}{\texttt{lenstronomy}\xspace}

\newcommand{\amplification}{\bm{\mathcal{A}}}

\newcommand\e[1]{_{\mathrm{#1}}}
\newcommand\h[1]{^{\mathrm{#1}}}

\newcommand{\dd}{\mathrm{d}}

\newcommand{\ddf}[3][]{\frac{\dd^{#1} #2}{\dd {#3}^{#1}}}



%% file: los_slacs_ii_oja_accepted.bib
@ARTICLE{2019MNRAS.483.2671B,
       author = {{Breton}, Michel-Andr{\`e}s and {Rasera}, Yann and {Taruya}, Atsushi and {Lacombe}, Osmin and {Saga}, Shohei},
        title = "{Imprints of relativistic effects on the asymmetry of the halo cross-correlation function: from linear to non-linear scales}",
      journal = {\mnras},
         year = 2019,
        month = feb,
       volume = {483},
       number = {2},
        pages = {2671-2696},
          doi = {10.1093/mnras/sty3206},
archivePrefix = {arXiv},
       eprint = {1803.04294},
 primaryClass = {astro-ph.CO},
       adsurl = {https://ui.adsabs.harvard.edu/abs/2019MNRAS.483.2671B}
}

@article{Kochanek_2021,
   title={Overconstrained models of time delay lenses redux: how the angular tail wags the radial dog},
   volume={501},
   ISSN={1365-2966},
   url={http://dx.doi.org/10.1093/mnras/staa4033},
   DOI={10.1093/mnras/staa4033},
   number={4},
   journal={Monthly Notices of the Royal Astronomical Society},
   publisher={Oxford University Press (OUP)},
   author={Kochanek, C S},
   year={2021},
   month=Jan, pages={5021–5028} 
}

@article{Johnson2025,
   title={Line-of-sight effects on double source plane lenses},
   volume={2025},
   ISSN={1475-7516},
   url={http://dx.doi.org/10.1088/1475-7516/2025/08/067},
   DOI={10.1088/1475-7516/2025/08/067},
   number={08},
   journal={Journal of Cosmology and Astroparticle Physics},
   publisher={IOP Publishing},
   author={Johnson, Daniel and Collett, Thomas and Li, Tian and Fleury, Pierre},
   year={2025},
   month=aug, pages={067} 
}

@article{McCully2017,
	doi = {10.3847/1538-4357/836/1/141},
	url = {https://doi.org/10.3847%2F1538-4357%2F836%2F1%2F141},
	year = 2017,
	month = {02},
	publisher = {American Astronomical Society},
	volume = {836},
	number = {1},
	pages = {141},
	author = {Curtis McCully and Charles R. Keeton and Kenneth C. Wong and Ann I. Zabludoff},
	title = {Quantifying Environmental and Line-of-sight Effects in Models of Strong Gravitational Lens Systems},
    eprint={1601.05417},
	journal = {The Astrophysical Journal}
}

@ARTICLE{Jaroszynski2014,
       author = {{Jaroszynski}, M. and {Kostrzewa-Rutkowska}, Z.},
        title = "{The influence of the matter along the line of sight and in the lens environment on the strong gravitational lensing}",
      journal = {Monthly Notices of the Royal Astronomical Society},
         year = 2014,
        month = 04,
       volume = {439},
       number = {3},
        pages = {2432-2441},
          doi = {10.1093/mnras/stu096},
archivePrefix = {arXiv},
       eprint = {1401.4108},
 primaryClass = {astro-ph.CO},
       adsurl = {https://ui.adsabs.harvard.edu/abs/2014MNRAS.439.2432J}
}

@article{Birrer:2016xku,
	title        = {{Line-of-sight effects in strong lensing: Putting theory into practice}},
	author       = {Birrer, Simon and Welschen, Cyril and Amara, Adam and Refregier, Alexandre},
	year         = 2017,
	journal      = {Journal of Cosmology and Astroparticle Physics},
	volume       = {04},
	pages        = {049},
	doi          = {10.1088/1475-7516/2017/04/049},
	eprint       = {1610.01599},
	archiveprefix = {arXiv},
	primaryclass = {astro-ph.CO}
}

@article{Birrer:2017sge,
	title        = {{Cosmic Shear with Einstein Rings}},
	author       = {Birrer, S. and Refregier, A. and Amara, A.},
	year         = 2018,
	journal      = {The Astrophysical Journal},
	volume       = 852,
	number       = 1,
	pages        = {L14},
	doi          = {10.3847/2041-8213/aaa1de},
	eprint       = {1710.01303},
	archiveprefix = {arXiv},
	primaryclass = {astro-ph.CO},
	slaccitation = {%%CITATION = ARXIV:1710.01303;%%}
}

@article{Birrer:2018xgm,
	title        = {lenstronomy: Multi-purpose gravitational lens modelling software package},
	author       = {Simon Birrer and Adam Amara},
	year         = 2018,
	month        = 3,
	journal      = {Physics of the Dark Universe},
	volume       = 22,
	pages        = {189--201},
	doi          = {https://doi.org/10.1016/j.dark.2018.11.002},
	issn         = {2212-6864}
}

@article{Birrer2021,
	title        = {lenstronomy II: A gravitational lensing software ecosystem},
	author       = {Simon Birrer and Anowar J. Shajib and Daniel Gilman and Aymeric Galan and Jelle Aalbers and Martin Millon and Robert Morgan and Giulia Pagano and Ji Won Park and Luca Teodori and Nicolas Tessore and Madison Ueland and Lyne Van de Vyvere and Sebastian Wagner-Carena and Ewoud Wempe and Lilan Yang and Xuheng Ding and Thomas Schmidt and Dominique Sluse and Ming Zhang and Adam Amara},
	year         = 2021,
	journal      = {Journal of Open Source Software},
	publisher    = {The Open Journal},
	volume       = 6,
	number       = 62,
	pages        = 3283,
	doi          = {10.21105/joss.03283},
	url          = {https://doi.org/10.21105/joss.03283}
}

@article{Bolton2006,
	title        = {{The Sloan Lens ACS Survey. I. A Large Spectroscopically Selected Sample of Massive Early-Type Lens Galaxies}},
	author       = {{Bolton}, A. S. and {Burles}, S.t and {Koopmans}, L. V.~E. and {Treu}, T. and {Moustakas}, L. A.},
	year         = 2006,
	month        = feb,
	journal      = {The Astrophysical Journal},
	volume       = 638,
	number       = 2,
	pages        = {703--724},
	doi          = {10.1086/498884},
	archiveprefix = {arXiv},
	eprint       = {astro-ph/0511453},
	primaryclass = {astro-ph},
	adsurl       = {https://ui.adsabs.harvard.edu/abs/2006ApJ...638..703B}
}

@article{Bolton2008,
	title        = {{The Sloan Lens ACS Survey. V. The Full ACS Strong-Lens Sample}},
	author       = {{Bolton}, A. S. and others},
	year         = 2008,
	month        = aug,
	journal      = {The Astrophysical Journal},
	volume       = 682,
	number       = 2,
	pages        = {964--984},
	doi          = {10.1086/589327},
	archiveprefix = {arXiv},
	eprint       = {0805.1931},
	primaryclass = {astro-ph},
	adsurl       = {https://ui.adsabs.harvard.edu/abs/2008ApJ...682..964B}
}

@article{Casey:2022amu,
	title        = {{COSMOS-Web: An Overview of the JWST Cosmic Origins Survey}},
	author       = {Casey, C. M. and others},
	year         = 2023,
	journal      = {The Astrophysical Journal},
	volume       = 954,
	number       = 1,
	pages        = 31,
	doi          = {10.3847/1538-4357/acc2bc},
	eprint       = {2211.07865},
	archiveprefix = {arXiv},
	primaryclass = {astro-ph.GA}
}

@article{Etherington:2023yyh,
    author = "Etherington, Amy and others",
    title = "{Strong gravitational lensing\textquoteright{}s \textquoteleft{}external shear\textquoteright{} is not shear}",
    eprint = "2301.05244",
    archivePrefix = "arXiv",
    primaryClass = "astro-ph.CO",
    doi = "10.1093/mnras/stae1375",
    journal = "Monthly Notices of the Royal Astronomical Society",
    volume = "531",
    number = "3",
    pages = "3684--3697",
    year = "2024"
}

@article{Falco1985,
	title        = {{On model-dependent bounds on H 0 from gravitational images : application to Q 0957+561 A, B.}},
	author       = {{Falco}, E.~E. and {Gorenstein}, M.~V. and {Shapiro}, I.~I.},
	year         = 1985,
	month        = feb,
	journal      = {The Astrophysical Journal Letters},
	volume       = 289,
	pages        = {L1-L4},
	doi          = {10.1086/184422},
	adsurl       = {https://ui.adsabs.harvard.edu/abs/1985ApJ...289L...1F}
}

@article{Fleury:2021tke,
	title        = {{Line-of-sight effects in strong gravitational lensing}},
	author       = {Fleury, P. and Larena, J. and Uzan, J.-P.},
	year         = 2021,
	journal      = {Journal of Cosmology and Astroparticle Physics},
	volume       = {08},
	pages        = {024},
	doi          = {10.1088/1475-7516/2021/08/024},
	eprint       = {2104.08883},
	archiveprefix = {arXiv},
	primaryclass = {astro-ph.CO},
	reportnumber = {IFT-UAM/CSIC-21-39}
}

@article{ForemanMackey2013,
	title        = {{emcee: The MCMC Hammer}},
	author       = {{Foreman-Mackey}, D. and {Hogg}, D. W. and {Lang}, D. and {Goodman}, J.},
	year         = 2013,
	journal      = {Publications of the Astronomical Society of the Pacific},
	volume       = 125,
	number       = 925,
	pages        = 306,
	doi          = {10.1086/670067},
	archiveprefix = {arXiv},
	eprint       = {1202.3665},
	primaryclass = {astro-ph.IM},
	adsurl       = {https://ui.adsabs.harvard.edu/abs/2013PASP..125..306F}
}

@article{GoodmanWeare,
	title        = {{Ensemble samplers with affine invariance}},
	author       = {{Goodman}, J. and {Weare}, J.},
	year         = 2010,
	month        = jan,
	journal      = {Communications in Applied Mathematics and Computational Science},
	volume       = 5,
	number       = 1,
	pages        = {65--80},
	doi          = {10.2140/camcos.2010.5.65},
	adsurl       = {https://ui.adsabs.harvard.edu/abs/2010CAMCS...5...65G}
}

@article{Hogg:2022ycw,
	title        = {{Measuring line-of-sight shear with Einstein rings: a proof of concept}},
	author       = {Hogg, N. B. and Fleury, P. and Larena, J. and Martinelli, M.},
	year         = 2023,
	journal      = {Monthly Notices of the Royal Astronomical Society},
	volume       = 520,
	number       = 4,
	pages        = {5982--6000},
	doi          = {10.1093/mnras/stad512},
	eprint       = {2210.07210},
	archiveprefix = {arXiv},
	primaryclass = {astro-ph.CO}
}

@article{Rasera:2021mvk,
	title        = {{The RayGalGroupSims cosmological simulation suite for the study of relativistic effects: An application to lensing-matter clustering statistics}},
	author       = {Rasera, Y. and others},
	year         = 2022,
	journal      = {Astronomy \& Astrophysics},
	volume       = 661,
	pages        = {A90},
	doi          = {10.1051/0004-6361/202141908},
	eprint       = {2111.08745},
	archiveprefix = {arXiv},
	primaryclass = {astro-ph.CO}
}

@article{Schneider:2013wga,
	title        = {{Source-position transformation -- an approximate invariance in strong gravitational lensing}},
	author       = {Schneider, Peter and Sluse, Dominique},
	year         = 2014,
	journal      = {Astronomy \& Astrophysics},
	volume       = 564,
	pages        = {A103},
	doi          = {10.1051/0004-6361/201322106},
	eprint       = {1306.4675},
	archiveprefix = {arXiv},
	primaryclass = {astro-ph.CO}
}

@article{Shajib2021,
	title        = {{Dark matter haloes of massive elliptical galaxies at z {\ensuremath{\sim}} 0.2 are well described by the Navarro-Frenk-White profile}},
	author       = {{Shajib}, A. J. and {Treu}, T. and {Birrer}, S. and {Sonnenfeld}, A.},
	year         = 2021,
	month        = may,
	journal      = {Monthly Notices of the Royal Astronomical Society},
	volume       = 503,
	number       = 2,
	pages        = {2380--2405},
	doi          = {10.1093/mnras/stab536},
	archiveprefix = {arXiv},
	eprint       = {2008.11724},
	primaryclass = {astro-ph.GA},
	adsurl       = {https://ui.adsabs.harvard.edu/abs/2021MNRAS.503.2380S}
}

@article{vanderVelden2020,
	title        = {{CMasher: Scientific colormaps for making accessible, informative and 'cmashing' plots}},
	author       = {van der Velden, Ellert},
	year         = 2020,
	month        = feb,
	journal      = {The Journal of Open Source Software},
	volume       = 5,
	number       = 46,
	pages        = 2004,
	doi          = {10.21105/joss.02004},
	eid          = 2004,
	archiveprefix = {arXiv},
	eprint       = {2003.01069},
	primaryclass = {eess.IV},
	adsurl       = {https://ui.adsabs.harvard.edu/abs/2020JOSS....5.2004V}
}

@article{Euclid:2024jyk,
    author = "Acevedo Barroso, J. A. and others",
    collaboration = "Euclid",
    title = "{Euclid: The Early Release Observations Lens Search Experiment}",
    eprint = "2408.06217",
    archivePrefix = "arXiv",
    primaryClass = "astro-ph.GA",
    month = "8",
    year = "2024"
}

@INPROCEEDINGS{Avila2015,
       author = {{Avila}, R.~J. and {Hack}, W. and {Cara}, M. and {Borncamp}, D. and {Mack}, J. and {Smith}, L. and {Ubeda}, L.},
        title = "{DrizzlePac 2.0 - Introducing New Features}",
    booktitle = {Astronomical Data Analysis Software an Systems XXIV (ADASS XXIV)},
         year = 2015,
       editor = {{Taylor}, A.~R. and {Rosolowsky}, E.},
       series = {Astronomical Society of the Pacific Conference Series},
       volume = {495},
        month = sep,
        pages = {281},
          doi = {10.48550/arXiv.1411.5605},
archivePrefix = {arXiv},
       eprint = {1411.5605},
 primaryClass = {astro-ph.IM},
       adsurl = {https://ui.adsabs.harvard.edu/abs/2015ASPC..495..281A}
}

@inproceedings{Krist2011,
author = {John E. Krist and Richard N. Hook and Felix Stoehr},
title = {{20 years of Hubble Space Telescope optical modeling using Tiny Tim}},
volume = {8127},
booktitle = {Optical Modeling and Performance Predictions V},
editor = {Mark A. Kahan},
organization = {International Society for Optics and Photonics},
publisher = {SPIE},
pages = {81270J},
year = {2011},
doi = {10.1117/12.892762},
URL = {https://doi.org/10.1117/12.892762}
}

@article{Tessore:2015baa,
    author = "Tessore, Nicolas and Metcalf, R. Benton",
    title = "{The elliptical power law profile lens}",
    eprint = "1507.01819",
    archivePrefix = "arXiv",
    primaryClass = "astro-ph.CO",
    doi = "10.1051/0004-6361/201526773",
    journal = "Astronomy \& Astrophysics",
    volume = "580",
    pages = "A79",
    year = "2015"
}

@ARTICLE{Sersic1963,
       author = {{S{\'e}rsic}, J.~L.},
        title = "{Influence of the atmospheric and instrumental dispersion on the brightness distribution in a galaxy}",
      journal = {Boletin de la Asociacion Argentina de Astronomia La Plata Argentina},
         year = 1963,
        month = feb,
       volume = {6},
        pages = {41-43},
       adsurl = {https://ui.adsabs.harvard.edu/abs/1963BAAA....6...41S}
}

@article{Refregier:2001fd,
    author = "Refregier, Alexandre",
    title = "{Shapelets: I. a method for image analysis}",
    eprint = "astro-ph/0105178",
    archivePrefix = "arXiv",
    doi = "10.1046/j.1365-8711.2003.05901.x",
    journal = "Monthly Notices of the Royal Astronomical Society",
    volume = "338",
    pages = "35",
    year = "2003"
}

@ARTICLE{Tan2024,
	author = {{Tan}, Chin Yi and {Shajib}, Anowar J. and {Birrer}, Simon and {Sonnenfeld}, Alessandro and {Treu}, Tommaso and {Wells}, Patrick and {Williams}, Devon and {Buckley-Geer}, Elizabeth J. and {Drlica-Wagner}, Alex and {Frieman}, Joshua},
	title = "{Project Dinos I: A joint lensing-dynamics constraint on the deviation from the power law in the mass profile of massive ellipticals}",
	journal = {Monthly Notices of the Royal Astronomical Society},
	year = 2024,
	month = may,
	volume = {530},
	number = {2},
	pages = {1474-1505},
	doi = {10.1093/mnras/stae884},
	archivePrefix = {arXiv},
	eprint = {2311.09307},
	primaryClass = {astro-ph.GA},
	adsurl = {https://ui.adsabs.harvard.edu/abs/2024MNRAS.530.1474T}
}

@book{Sprent1998,
	title        = "{Data Driven Statistical Methods}",
	author       = {P. Sprent},
	year         = 1998,
	publisher    = {Chapman \& Hall}
}

@ARTICLE{VandeVyvere2021,
	author = {{Van de Vyvere}, Lyne and {Gomer}, Matthew R. and {Sluse}, Dominique and {Xu}, Dandan and {Birrer}, Simon and {Galan}, Aymeric and {Vernardos}, Georgios},
	title = "{TDCOSMO. VII. Boxyness/discyness in lensing galaxies: Detectability and impact on H$_{0}$}",
	journal = {Astronomy \& Astrophysics},
	year = 2022,
	month = mar,
	volume = {659},
	eid = {A127},
	pages = {A127},
	doi = {10.1051/0004-6361/202141551},
	archivePrefix = {arXiv},
	eprint = {2112.03932},
	primaryClass = {astro-ph.CO},
	adsurl = {https://ui.adsabs.harvard.edu/abs/2022A&A...659A.127V}
}

@article{Prat:2025ucy,
	author = "Prat, J. and Bacon, D.",
	title = "{Weak Gravitational Lensing}",
	eprint = "2501.07938",
	archivePrefix = "arXiv",
	primaryClass = "astro-ph.CO",
	month = "1",
	year = "2025"
}

@ARTICLE{Bender1988,
	author = {{Bender}, R. and {Doebereiner}, S. and {Moellenhoff}, C.},
	title = "{Isophote shapes of elliptical galaxies. I. The data.}",
	journal = {Astronomy \& Astrophysics Supplement},
	year = 1988,
	month = sep,
	volume = {74},
	pages = {385-426},
	adsurl = {https://ui.adsabs.harvard.edu/abs/1988A&AS...74..385B}
}

@ARTICLE{Shajib2024,
	author = {{Shajib}, A.~J. and {Vernardos}, G. and {Collett}, T.~E. and {Motta}, V. and {Sluse}, D. and {Williams}, L.~L.~R. and {Saha}, P. and {Birrer}, S. and {Spiniello}, C. and {Treu}, T.},
	title = "{Strong Lensing by Galaxies}",
	journal = {Space Science Reviews},
	year = 2024,
	month = dec,
	volume = {220},
	number = {8},
	eid = {87},
	pages = {87},
	doi = {10.1007/s11214-024-01105-x},
	archivePrefix = {arXiv},
	eprint = {2210.10790},
	primaryClass = {astro-ph.GA},
	adsurl = {https://ui.adsabs.harvard.edu/abs/2024SSRv..220...87S}
}

@BOOK{Sersic1968,
	author = {{S{\'e}rsic}, J.~L.},
	title = "{Atlas de Galaxias Australes}",
	year = 1968,
	adsurl = {https://ui.adsabs.harvard.edu/abs/1968adga.book.....S}
}

@article{Johnson:2024hvl,
	author = "Johnson, Daniel and Fleury, Pierre and Larena, Julien and Marchetti, Lucia",
	title = "{Foreground biases in strong gravitational lensing}",
	eprint = "2405.04194",
	archivePrefix = "arXiv",
	primaryClass = "astro-ph.CO",
	doi = "10.1088/1475-7516/2024/10/055",
	journal = "Journal of Cosmology and Astroparticle Physics",
	volume = "10",
	pages = "055",
	year = "2024"
}

@article{Hogg:2025wac,
    author = "Hogg, Natalie B. and Johnson, Daniel P. and Shajib, Anowar J. and Larena, Julien",
    title = "{Line-of-sight shear in SLACS strong lenses I: shear and mass model parametrisations}",
    eprint = "2501.16292",
    archivePrefix = "arXiv",
    primaryClass = "astro-ph.CO",
    doi = "10.33232/001c.164832",
    journal = "Open J. Astrophys.",
    volume = "9",
    pages = "164832",
    year = "2026"
}

@ARTICLE{Green2011,
       author = {{Green}, D.~A.},
        title = "{A colour scheme for the display of astronomical intensity images}",
      journal = {Bulletin of the Astronomical Society of India},
         year = 2011,
        month = jun,
       volume = {39},
        pages = {289-295},
          doi = {10.48550/arXiv.1108.5083},
archivePrefix = {arXiv},
       eprint = {1108.5083},
 primaryClass = {astro-ph.IM},
       adsurl = {https://ui.adsabs.harvard.edu/abs/2011BASI...39..289G}
}

@ARTICLE{Thyng2016,
       author = {{Thyng}, Kristen and {Greene}, Chad and {Hetland}, Robert and {Zimmerle}, Heather and {DiMarco}, Steven},
        title = "{True Colors of Oceanography: Guidelines for Effective and Accurate Colormap Selection}",
      journal = {Oceanography},
         year = 2016,
        month = sep,
       volume = {29},
       number = {3},
        pages = {9-13},
          doi = {10.5670/oceanog.2016.66},
       adsurl = {https://ui.adsabs.harvard.edu/abs/2016Ocgpy..29c...9T}
}

@article{Shajib:2025bho,
    author = "Shajib, Anowar J. and Nihal, Nafis Sadik and Tan, Chin Yi and Sahu, Vedant and Birrer, Simon and Treu, Tommaso and Frieman, Joshua",
    title = "{dolphin: A Fully Automated Forward-modeling Pipeline Powered by Artificial Intelligence for Galaxy-scale Strong Lenses}",
    eprint = "2503.22657",
    archivePrefix = "arXiv",
    primaryClass = "astro-ph.IM",
    doi = "10.3847/1538-4357/adf95c",
    journal = "The Astrophysical Journal",
    volume = "992",
    number = "1",
    pages = "40",
    year = "2025"
}

@INPROCEEDINGS{Eberhart1995,
  author={Eberhart, R. and Kennedy, J.},
  booktitle={MHS'95. Proceedings of the Sixth International Symposium on Micro Machine and Human Science}, 
  title={A new optimizer using particle swarm theory}, 
  year={1995},
  volume={},
  number={},
  pages={39-43},
  doi={10.1109/MHS.1995.494215}}

@article{Scognamiglio:2026phv,
    author = "Scognamiglio, Diana and others",
    title = "{An ultra-high-resolution map of (dark) matter}",
    eprint = "2601.17239",
    archivePrefix = "arXiv",
    primaryClass = "astro-ph.CO",
    doi = "10.1038/s41550-025-02763-9",
    journal = "Nature Astron.",
    volume = "10",
    number = "4",
    pages = "573--582",
    year = "2026"
}

@article{Kaiser:2000if,
    author = "Kaiser, Nick and Wilson, Gillian and Luppino, Gerard A.",
    title = "{Large scale cosmic shear measurements}",
    eprint = "astro-ph/0003338",
    archivePrefix = "arXiv",
    month = "3",
    year = "2000"
}

@ARTICLE{vanWaerbeke2000,
       author = {{Van Waerbeke}, L. and {Mellier}, Y. and {Erben}, T. and {Cuillandre}, J.~C. and {Bernardeau}, F. and {Maoli}, R. and {Bertin}, E. and {McCracken}, H.~J. and {Le F{\`e}vre}, O. and {Fort}, B. and {Dantel-Fort}, M. and {Jain}, B. and {Schneider}, P.},
        title = "{Detection of correlated galaxy ellipticities from CFHT data: first evidence for gravitational lensing by large-scale structures}",
      journal = {Astronomy \& Astrophysics},
         year = 2000,
        month = jun,
       volume = {358},
        pages = {30-44},
          doi = {10.48550/arXiv.astro-ph/0002500},
archivePrefix = {arXiv},
       eprint = {astro-ph/0002500},
 primaryClass = {astro-ph},
       adsurl = {https://ui.adsabs.harvard.edu/abs/2000A&A...358...30V}
}

@article{Wittman:2000tc,
    author = "Wittman, David M. and Tyson, J. Anthony and Kirkman, David and Dell'Antonio, Ian and Bernstein, Gary",
    title = "{Detection of weak gravitational lensing distortions of distant galaxies by cosmic dark matter at large scales}",
    eprint = "astro-ph/0003014",
    archivePrefix = "arXiv",
    doi = "10.1038/35012001",
    journal = "Nature",
    volume = "405",
    pages = "143--149",
    year = "2000"
}

@article{Bacon:2000sy,
    author = "Bacon, David J. and Refregier, Alexandre R. and Ellis, Richard S.",
    title = "{Detection of weak gravitational lensing by large-scale structure}",
    eprint = "astro-ph/0003008",
    archivePrefix = "arXiv",
    doi = "10.1046/j.1365-8711.2000.03851.x",
    journal = "Monthly Notices of the Royal Astronomical Society",
    volume = "318",
    pages = "625",
    year = "2000"
}

@ARTICLE{Lin2012,
       author = {{Lin}, Huan and {Dodelson}, Scott and {Seo}, Hee-Jong and {Soares-Santos}, Marcelle and {Annis}, James and {Hao}, Jiangang and {Johnston}, David and {Kubo}, Jeffrey M. and {Reis}, Ribamar R.~R. and {Simet}, Melanie},
        title = "{The SDSS Co-add: Cosmic Shear Measurement}",
      journal = {\apj},
         year = 2012,
        month = dec,
       volume = {761},
       number = {1},
          eid = {15},
        pages = {15},
          doi = {10.1088/0004-637X/761/1/15},
archivePrefix = {arXiv},
       eprint = {1111.6622},
 primaryClass = {astro-ph.CO},
       adsurl = {https://ui.adsabs.harvard.edu/abs/2012ApJ...761...15L}
}

@article{Kuijken:2015vca,
    author = "Kuijken, Konrad and others",
    title = "{Gravitational Lensing Analysis of the Kilo Degree Survey}",
    eprint = "1507.00738",
    archivePrefix = "arXiv",
    primaryClass = "astro-ph.CO",
    doi = "10.1093/mnras/stv2140",
    journal = "Monthly Notices of the Royal Astronomical Society",
    volume = "454",
    number = "4",
    pages = "3500--3532",
    year = "2015"
}

@ARTICLE{Kilbinger2013,
       author = {{Kilbinger}, Martin and {Fu}, Liping and {Heymans}, Catherine and {Simpson}, Fergus and {Benjamin}, Jonathan and {Erben}, Thomas and {Harnois-D{\'e}raps}, Joachim and {Hoekstra}, Henk and {Hildebrandt}, Hendrik and {Kitching}, Thomas D. and {Mellier}, Yannick and {Miller}, Lance and {Van Waerbeke}, Ludovic and {Benabed}, Karim and {Bonnett}, Christopher and {Coupon}, Jean and {Hudson}, Michael J. and {Kuijken}, Konrad and {Rowe}, Barnaby and {Schrabback}, Tim and {Semboloni}, Elisabetta and {Vafaei}, Sanaz and {Velander}, Malin},
        title = "{CFHTLenS: combined probe cosmological model comparison using 2D weak gravitational lensing}",
      journal = {Montly Notices of the Royal Astronomical Society},
         year = 2013,
        month = apr,
       volume = {430},
       number = {3},
        pages = {2200-2220},
          doi = {10.1093/mnras/stt041},
archivePrefix = {arXiv},
       eprint = {1212.3338},
 primaryClass = {astro-ph.CO},
       adsurl = {https://ui.adsabs.harvard.edu/abs/2013MNRAS.430.2200K}
}

@article{Secco2021,
    author = "Secco, L. F. and others",
    collaboration = "DES",
    title = "{Dark Energy Survey Year 3 results: Cosmology from cosmic shear and robustness to modeling uncertainty}",
    eprint = "2105.13544",
    archivePrefix = "arXiv",
    primaryClass = "astro-ph.CO",
    reportNumber = "FERMILAB-PUB-21-253-AE, DES-2019-0480",
    doi = "10.1103/PhysRevD.105.023515",
    journal = "Physical Review D",
    volume = "105",
    number = "2",
    pages = "023515",
    year = "2022"
}

@article{Kilbinger:2014cea,
    author = "Kilbinger, Martin",
    title = "{Cosmology with cosmic shear observations: a review}",
    eprint = "1411.0115",
    archivePrefix = "arXiv",
    primaryClass = "astro-ph.CO",
    doi = "10.1088/0034-4885/78/8/086901",
    journal = "Reports on Progress in Physics",
    volume = "78",
    pages = "086901",
    year = "2015"
}

@article{Nightingale:2025mlk,
    author = "Nightingale, James and others",
    title = "{The COSMOS-Web Lens Survey (COWLS) I: Discovery of {\ensuremath{>}}100 high redshift strong lenses in contiguous JWST imaging}",
    eprint = "2503.08777",
    archivePrefix = "arXiv",
    primaryClass = "astro-ph.GA",
    doi = "10.1093/mnras/staf1253",
    journal = "Monthly Notices of the Royal Astronomical Society",
    volume = "543",
    number = "1",
    pages = "203--222",
    year = "2025"
}

@article{Mahler:2025ajl,
    author = "Mahler, Guillaume and others",
    title = "{The COSMOS-Web Lens Survey (COWLS) II: depth, resolution, and NIR coverage from JWST reveal 17 spectacular lenses}",
    eprint = "2503.08782",
    archivePrefix = "arXiv",
    primaryClass = "astro-ph.GA",
    doi = "10.1093/mnrasl/slaf088",
    journal = "Monthly Notices of the Royal Astronomical Society: Letters",
    volume = "8",
    pages = "L14",
    year = "2025"
}

@article{Hogg:2025plt,
    author = "Hogg, Natalie B. and others",
    title = "{The COSMOS-Web Lens Survey (COWLS) III: forecasts versus data}",
    eprint = "2503.08785",
    archivePrefix = "arXiv",
    primaryClass = "astro-ph.GA",
    month = "3",
    year = "2025",
    journal = "Monthly Notices of the Royal Astronomical Society",
    volume = "544",
    number = "1",
    pages = "782--798"
}

@article{Euclid:2025rfv,
    author = "Walmsley, M. and others",
    collaboration = "Euclid",
    title = "{Euclid Quick Data Release (Q1): The Strong Lensing Discovery Engine A -- System overview and lens catalogue}",
    eprint = "2503.15324",
    archivePrefix = "arXiv",
    primaryClass = "astro-ph.GA",
    month = "3",
    year = "2025"
}

@article{Legin:2025abc,
    author = "Legin, Ronan and Stone, Connor and Adam, Alexandre and Barco, Gabriel Missael and Coogan, Adam and Malkin, Nikolay and Perreault-Levasseur, Laurence and Hezaveh, Yashar",
    title = "{Mind the Information Gap: Unveiling Detailed Morphologies of z 0.5-1.0 Galaxies with SLACS Strong Lenses and Data-Driven Analysis}",
    eprint = "2511.19595",
    archivePrefix = "arXiv",
    primaryClass = "astro-ph.GA",
    month = "11",
    year = "2025"
}

@article{PerreaultLevasseur:2017ltk,
    author = "Perreault Levasseur, Laurence and Hezaveh, Yashar D. and Wechsler, Risa H.",
    title = "{Uncertainties in Parameters Estimated with Neural Networks: Application to Strong Gravitational Lensing}",
    eprint = "1708.08843",
    archivePrefix = "arXiv",
    primaryClass = "astro-ph.CO",
    doi = "10.3847/2041-8213/aa9704",
    journal = "The Astrophysical Journal Letters",
    volume = "850",
    number = "1",
    pages = "L7",
    year = "2017"
}

@article{Wagner-Carena:2020yun,
    author = "Wagner-Carena, Sebastian and Park, Ji Won and Birrer, Simon and Marshall, Philip J. and Roodman, Aaron and Wechsler, Risa H.",
    collaboration = "LSST Dark Energy Science",
    title = "{Hierarchical Inference with Bayesian Neural Networks: An Application to Strong Gravitational Lensing}",
    eprint = "2010.13787",
    archivePrefix = "arXiv",
    primaryClass = "astro-ph.CO",
    doi = "10.3847/1538-4357/abdf59",
    journal = "The Astrophysical Journal",
    volume = "909",
    number = "2",
    pages = "187",
    year = "2021"
}

@article{Mukherjee:2018cdf,
    author = "Mukherjee, Sampath and others",
    title = "{SEAGLE {\textendash} I. A pipeline for simulating and modelling strong lenses from cosmological hydrodynamic simulations}",
    eprint = "1802.06629",
    archivePrefix = "arXiv",
    primaryClass = "astro-ph.CO",
    doi = "10.1093/mnras/sty1741",
    journal = "Monthly Notices of the Royal Astronomical Society",
    volume = "479",
    number = "3",
    pages = "4108--4125",
    year = "2018"
}

@article{Schaye:2025xuv,
    author = "Schaye, Joop and others",
    title = "{The COLIBRE project: cosmological hydrodynamical simulations of galaxy formation and evolution}",
    eprint = "2508.21126",
    archivePrefix = "arXiv",
    primaryClass = "astro-ph.GA",
    month = "8",
    year = "2025"
}

@ARTICLE{DECV2025,
       author = {{Anbajagane}, D. and {Chang}, C. and {Drlica-Wagner}, A. and {Tan}, C.~Y. and {Adamow}, M. and {Gruendl}, R.~A. and {Secco}, L.~F. and {Zhang}, Z. and {Becker}, M.~R. and {Ferguson}, P.~S. and {Chicoine}, N. and {Herron}, K. and {Alarcon}, A. and {Teixeira}, R. and {Suson}, D. and {Shajib}, A.~J. and {Frieman}, J.~A. and {Alsina}, A.~N. and {Amon}, A. and {Andrade-Oliveira}, F. and {Blazek}, J. and {Bom}, C.~R. and {Camacho}, H. and {Carballo-Bello}, J.~A. and {Carnero Rosell}, A. and {Cawthon}, R. and {Cerny}, W. and {Choi}, A. and {Choi}, Y. and {Dodelson}, S. and {Doux}, C. and {Eckert}, K. and {Elvin-Poole}, J. and {Esteves}, J. and {Gatti}, M. and {Giannini}, G. and {Gruen}, D. and {Hartley}, W.~G. and {Herner}, K. and {Huff}, E.~M. and {Jain}, B. and {James}, D.~J. and {Jarvis}, M. and {Krause}, E. and {Kuropatkin}, N. and {Mart{\'\i}nez-V{\'a}zquez}, C.~E. and {Massana}, P. and {Mau}, S. and {McCullough}, J. and {Medina}, G.~E. and {Mutlu-Pakdil}, B. and {Myles}, J. and {Navabi}, M. and {No{\"e}l}, N.~E.~D. and {Pace}, A.~B. and {Pandey}, S. and {Porredon}, A. and {Prat}, J. and {Raveri}, M. and {Riley}, A.~H. and {Rykoff}, E.~S. and {Sakowska}, J.~D. and {Samuroff}, S. and {Sanchez-Cid}, D. and {Sand}, D.~J. and {Santana-Silva}, L. and {Sevilla-Noarbe}, I. and {Shin}, T. and {Soares-Santos}, M. and {Stringfellow}, G.~S. and {To}, C. and {Tollerud}, E.~J. and {Tong}, A. and {Troxel}, M.~A. and {Vivas}, A.~K. and {Yamamoto}, M. and {Yanny}, B. and {Yin}, B. and {Zenteno}, A. and {Zhang}, Y. and {Zuntz}, J.},
        title = "{The Dark Energy Camera All Data Everywhere cosmic shear project V: Constraints on cosmology and astrophysics from 270 million galaxies across 13,000 deg$^2$ of the sky}",
      journal = {arXiv e-prints},
         year = 2025,
        month = sep,
          eid = {arXiv:2509.03582},
        pages = {arXiv:2509.03582},
          doi = {10.48550/arXiv.2509.03582},
archivePrefix = {arXiv},
       eprint = {2509.03582},
 primaryClass = {astro-ph.CO},
       adsurl = {https://ui.adsabs.harvard.edu/abs/2025arXiv250903582A}
}

@article{DES:2021vln,
    author = "Secco, L. F. and others",
    collaboration = "DES",
    title = "{Dark Energy Survey Year 3 results: Cosmology from cosmic shear and robustness to modeling uncertainty}",
    eprint = "2105.13544",
    archivePrefix = "arXiv",
    primaryClass = "astro-ph.CO",
    reportNumber = "FERMILAB-PUB-21-253-AE, DES-2019-0480",
    doi = "10.1103/PhysRevD.105.023515",
    journal = "Physical Review D",
    volume = "105",
    number = "2",
    pages = "023515",
    year = "2022"
}

@article{DES:2021bvc,
    author = "Amon, A. and others",
    collaboration = "DES",
    title = "{Dark Energy Survey Year 3 results: Cosmology from cosmic shear and robustness to data calibration}",
    eprint = "2105.13543",
    archivePrefix = "arXiv",
    primaryClass = "astro-ph.CO",
    reportNumber = "FERMILAB-PUB-21-250-AE, DES-2019-0479",
    doi = "10.1103/PhysRevD.105.023514",
    journal = "Physical Review D",
    volume = "105",
    number = "2",
    pages = "023514",
    year = "2022"
}

@article{Shajib:2019crn,
    author = "Shajib, Anowar J.",
    title = "{Unified lensing and kinematic analysis for any elliptical mass profile}",
    eprint = "1906.08263",
    archivePrefix = "arXiv",
    primaryClass = "astro-ph.CO",
    doi = "10.1093/mnras/stz1796",
    journal = "Monthly Notices of the Royal Astronomical Society",
    volume = "488",
    number = "1",
    pages = "1387--1400",
    year = "2019"
}

@article{Galan:2020mnn,
    author = "Galan, A. and Peel, A. and Joseph, R. and Courbin, F. and Starck, J. -L",
    title = "{SLITronomy: towards a fully wavelet-based strong lensing inversion technique}",
    eprint = "2012.02802",
    archivePrefix = "arXiv",
    primaryClass = "astro-ph.GA",
    doi = "10.1051/0004-6361/202039363",
    journal = "Astronomy \& Astrophysics",
    volume = "647",
    pages = "A176",
    year = "2021"
}

@ARTICLE{Vernardos2022,
       author = {{Vernardos}, G. and {Koopmans}, L.~V.~E.},
        title = "{The very knotty lenser: Exploring the role of regularization in source and potential reconstructions using Gaussian process regression}",
      journal = {Monthly Notices of the Royal Astronomical Society},
         year = 2022,
        month = oct,
       volume = {516},
       number = {1},
        pages = {1347-1372},
          doi = {10.1093/mnras/stac1924},
archivePrefix = {arXiv},
       eprint = {2202.09378},
 primaryClass = {astro-ph.GA},
       adsurl = {https://ui.adsabs.harvard.edu/abs/2022MNRAS.516.1347V}
}

@article{Fleury:2026pco,
    author = "Fleury, Pierre and Johnson, Daniel and Duboscq, Th{\'e}o and Hogg, Natalie B. and Larena, Julien",
    title = "{Cosmology with the line-of-sight shear of strong gravitational lenses}",
    eprint = "2603.03441",
    archivePrefix = "arXiv",
    primaryClass = "astro-ph.CO",
    month = "3",
    year = "2026"
}
